\documentclass[a4paper,fleqn]{cas-dc}

\usepackage{tcolorbox}
\usepackage{subcaption}
 \usepackage{tabularray}

\usepackage{tikz}
\usepackage{xcolor,colortbl}
\usepackage{color}
 \usepackage{graphicx}
\usepackage[linesnumbered, vlined, ruled,algo2e]{algorithm2e}
\usepackage{soul}
 \usepackage{hyperref}
 \hypersetup{
	colorlinks,%
	citecolor=black,%
	filecolor=black,%
	linkcolor=black,%
	urlcolor=black,%
pdftitle={Adaptive Indexing for Approximate Query Processing in Exploratory Data Analysis},
pdfsubject={Incremental Indexing for Approximate Aggregation Queries in Visual Analytics},
pdfkeywords={Approximate aggregations, Aggregation Queries, Progressive incremental indexing, Big Data progressive visualization tools, Interactive visual sampling, Real-time data analysis, Interactive indexes, Learning indexes,  Information visualization, Error-bounded queries, Big Raw Data, HCI, RawVis, In-situ Query Processing, Human-Data interaction.},
pdfauthor={Stavros Maroulis, Nikos Bikakis,  Vassilis Stamatopoulos, George Papastefanatos}
}
 
\definecolor{myboxcolor}{RGB}{218, 233, 247}

\graphicspath{{figures/}{plots/}}

\definecolor{dark-gray}{gray}{0.2}

\newcommand\dunderline[3][-1pt]{{%
  \sbox0{#3}%
  \ooalign{\copy0\cr\rule[\dimexpr#1-#2\relax]{\wd0}{#2}}}}

\newcommand{\silence}[1]{}

\newcommand{\eat}[1]{}

\newcommand{\hidefornow}[1]{}        

\newcommand{\tofinish}[1]{}        

 \renewcommand{\O}{\ensuremath\mathcal{O}}
  \newcommand{\F}{\ensuremath\mathcal{F}}
   \newcommand{\A}{\ensuremath\mathcal{A}}
  \newcommand{\M}{\ensuremath\mathcal{M}}
  
  \newcommand{\C}{\ensuremath\mathcal{C}}

  \newcommand{\R}{\ensuremath\mathcal{R}}
  
 \renewcommand{\S}{\ensuremath\mathcal{S}}

 \newcommand{\T}{\ensuremath\mathcal{T}}
 \renewcommand{\L}{\ensuremath\mathcal{L}}

\renewcommand{\c}{\ensuremath\mathbf{c}}

\newcommand{\ind}{\mbox{VALINOR-A}\xspace}
\newcommand{\valinor}{\mbox{VALINOR}\xspace}

\newcommand{\val}{\mbox{VAL}\xspace}
\newcommand{\vals}{\mbox{VAL-S}\xspace}
\newcommand{\vala}{\mbox{VAL-A}\xspace}

\sethlcolor{green}

\newcommand*\circledFill[1]{\tikz[baseline=(char.base)]{
		\node[shape=circle,draw,inner sep=0.75pt, line width=1.0pt,text=white,fill=black](char) {\footnotesize{\textbf{#1}}};}}

\newcommand{\stitle}[1]{\bigskip\noindent\textbf{#1}} 
\newcommand{\sstitle}[1]{\vspace{0.2cm}\noindent\textit{#1}}

\newcounter{ex}
\newenvironment{myExample}
{\refstepcounter{ex}\vspace{8pt}\setlength{\leftskip}{0pt}\setlength{\rightskip}{0pt}\par\noindent\ignorespaces
   \textbf{Example~\theex.}}
{\vspace*{2pt}}

\begin{document}
\let\WriteBookmarks\relax
\def\floatpagepagefraction{1}
\def\textpagefraction{.001}

\title [mode = title]{Adaptive Indexing for Approximate Query Processing \linebreak in  Exploratory  Data Analysis}

\author[1]{Stavros Maroulis}
 [orcid=0000-0003-2816-4368]

\author[2,3]{Nikos Bikakis}
 [orcid=0000-0001-6859-1941]

\author[1,4]{Vassilis Stamatopoulos}
 [orcid=0000-0002-9044-796X]

\author[1]{George Papastefanatos}
 [orcid=0000-0002-9273-9843]

\affiliation[1]{organization={ATHENA Research Center},
                country={Greece}}

\affiliation[2]{organization={Hellenic Mediterranean University},
            city={Chania},
            country={Greece}}
\affiliation[3]{organization={Archimedes/Athena RC},
                 country={Greece}}

\affiliation[4]{organization={University of Ioannina},
                country={Greece}}

\begin{abstract}
Minimizing data-to-analysis time while enabling real-time interaction and efficient analytical computations on large datasets are fundamental objectives of contemporary exploratory systems.
Although some of the recent adaptive indexing and on-the-fly processing approaches address most of these needs, there are cases, where they do not always guarantee reliable performance.
Some examples of such cases include: exploring areas with a high density of objects;
executing the first exploratory queries or exploring previously unseen areas (where the index has not yet adapted sufficiently);
and working with very large data files on commodity hardware, such as low-specification laptops.
In such demanding cases, approximate and incremental techniques can be exploited to ensure
efficiency and scalability by allowing users to prioritize response time over result
accuracy, acknowledging that exact results are not always necessary.
Therefore, approximation mechanisms that enable smooth user interaction by defining the trade-off between accuracy and performance based on vital factors (e.g., task, preferences, available resources) are of great importance.
Considering the aforementioned, in this work, we present an
\textit{adaptive approximate query processing framework} for interactive on-the-fly analysis (with out a preprocessing phase)
 over large raw data. The core component of the framework is a
\textit{main-memory adaptive indexing scheme} (\ind) that interoperates with
\textit{user-driven sampling} and \textit{incremental aggregation computations}.
Additionally, an effective \textit{error-bounded approximation strategy} is designed  and integrated in the query processing  process.
We conduct extensive experiments using both real and synthetic datasets, demonstrating the efficiency and effectiveness of the proposed framework.
\end{abstract}


\maketitle



\section{Introduction}
\label{sec:intro}

A key objective in exploratory analysis is to minimize \textit{data-to-analysis time} while enabling \textit{real-time interactions} and \textit{efficient analytical computations} on very large data files.
In many cases, the development of approximate and incremental techniques is essential for addressing the aforementioned challenges, as they enable dynamic adjustments to system performance and result accuracy \cite{richer2024scalability,hal-04361344}.

The \textit{in-situ paradigm} is a common practice to minimize data-to-analysis time,
referring to on-the-fly data exploration and analysis of large raw data sets such as CSV or
JSON files \cite{MaroulisS22f, IS, Alagiannis2012, slalomvldbj19}.
In-situ techniques aim to bypass the overhead of fully loading and indexing data in a DBMS
(minimize the data-to-analysis time),  while offering efficient query evaluation.
In the context of in-situ data exploration,  previous works \cite{MaroulisS22f, IS} have
focused on building adaptive indexes that leverage the locality-based behavior of  data
exploration, by initially creating a "crude" version of the index around the initial area
of interest, and dynamically enriching and adapting the structure and its contents  (e.g.,
statistics) based on user interactions.

While in-situ setting ensures low  index initialization  cost (low data-to-analysis time), there are cases where poor query evaluation performance is observed.
Some examples of such cases include:
(1) when accessing areas with a \textit{high density of objects};
(2) \textit{during the initial queries} or when the user explores a \textit{previously unseen area}, where the index has not yet adapted sufficiently; and
(3) when exploring \textit{very large data files on commodity hardware}, such as low-specification laptop.

Meanwhile, in many interactive analytical tasks, users do not always \textit{require exact results}; instead response \textit{time is more crucial than result accuracy} \cite{hal-04361344, KimBPIMR15, 0001S20, ParkCM16, RahmanAKBKPR17, liu2013immens}.
For example, several visual analytic tasks, such as class or outlier analysis in scatterplots, pair-wise comparison of spatial areas on maps, often begin with {approximate}  insights, which can be used by the experts to quickly identify specific areas in the exploration space for further analysis.
The development of approximate and incremental techniques, will be able to guarantee efficient query evaluation and scalability in the challenging cases describe above, where poor performance is observed.

Although, \textit{approximate query processing} (AQP) \cite{MozafariN15,CormodeGHJ12,LiL18} is a long-studied problem in the areas of databases, data mining and information visualization; a gap exists when AQP is coupled with the in-situ setting.
While recent works in approximate query evaluation~\cite{AQP++, Liang21} combine {sampling with pre-aggregation} to improve confidence intervals and reduce query costs, these approaches rely on precomputed structures, which lack adaptability to evolving query workloads and user interactions. Moreover, their reliance on fully precomputed aggregates removes uncertainty but lacks adaptability to evolving query patterns. This rigidity limits their effectiveness in dynamic exploration scenarios where queries are unpredictable. To address this limitation, we integrate adaptive indexing over raw data files with incremental sampling. Instead of precomputing and storing exact aggregates, we leverage the samples collected during query evaluation to maintain \textit{approximate aggregates}. These aggregates can be reused in future queries, reducing unnecessary I/O while ensuring accuracy constraints are met.

\begin{figure*}[t]
 	\centering
    \vspace{6pt}
\includegraphics[width=\linewidth]{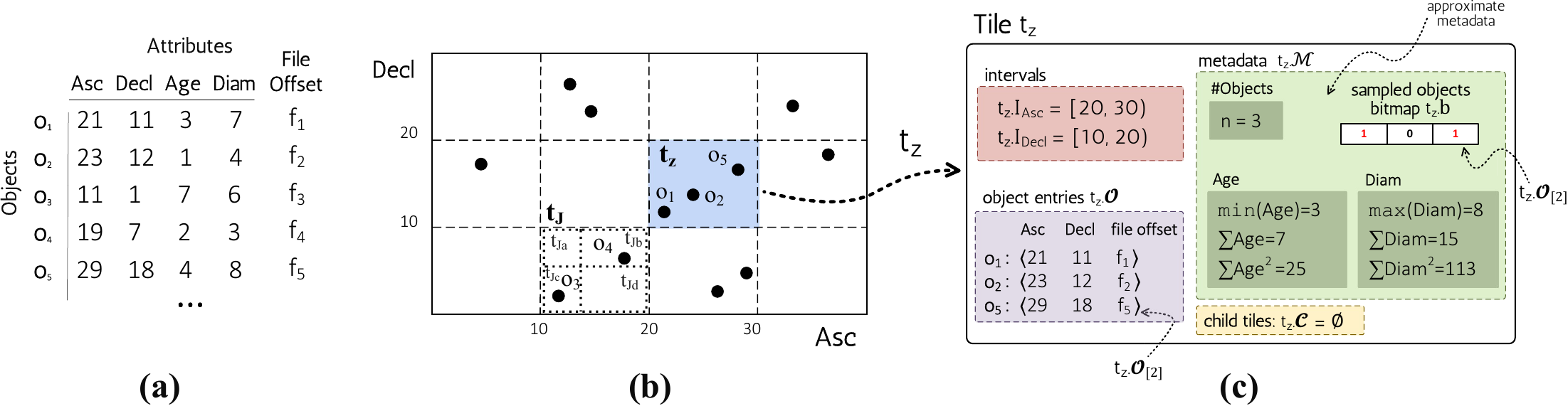 }
\vspace{6pt}
 \caption{Data and Index Example: (a) Data Objects; \hspace{1pt} (b) \ind Index; \hspace{1pt}  (c) \ind Tile }
 \label{fig:index}
  \end{figure*}

\stitle{Motivating Example.}
Consider an example where an astronomer uses they laptop to explore a very  large CSV file (e.g, Sloan Digital Sky Survey -- SDSS) containing celestial objects (e.g., stars) via a visual interface.
Each object is described by four attributes (Fig.~\ref{fig:index}a):
\textit{right ascension} (Asc) and \textit{declination} (Decl), that correspond to terrestrial longitude and  declination to geographic latitude;
\textit{age} (the age of the star in billion years);
and \textit{diameter} (the diameter in km).

The astronomer begins by inspecting a specific \textit{region of the sky} through a 2D visualization based on Asc and Decl, where objects are displayed alongside aggregate statistics, such as the \textit{average age} or \textit{maximum diameter}, computed for the current viewport.

As the astronomer interacts with the visualization (\textit{panning across sky regions or zooming in for detailed analysis}) the system dynamically updates both the displayed objects and their aggregate statistics. Initially, the focus is on broad regions to identify \textit{anomalous patterns}, such as clusters of \textit{exceptionally young stars} or \textit{unusually large diameters}. At this \textit{exploratory phase}, exact values are unnecessary, and approximations enable rapid insight generation.

Upon identifying an area of interest, the astronomer may require \textit{higher accuracy} to refine their analysis. To manage this trade-off, they can \textit{adjust the acceptable error bound}; allowing higher error for \textit{broad overviews} and anomaly detection while lowering it for \textit{precise measurements}. This approach enhances interactivity, ensuring a seamless transition from rapid exploration to detailed analysis.

\stitle{Basic Challenges.}
Some of the key challenges that arise in interactive data analysis over large  datasets, which do not fit into main memory, include:

\begin{itemize}

    \item \textbf{Low Data-to-Analysis Time.}
      Achieving a low data-to-analysis time requires eliminating a preprocessing
      phase; thus, tasks such as data organization, indexing, or analysis cannot
      be considered viable options.
        Therefore, efficiency and scalability must be achieved solely through on-the-fly processing.

      \item \textbf{Real-time Interaction.}
    Each user interaction (e.g., zooming, requesting statistics) triggers new queries that must be processed in real time.
    This is particularly challenging due to the high I/O operations cost
    associated with frequent raw file access.
    The challenge becomes even greater
    in areas with a \textit{high density of objects};
    during \textit{the first  queries}; and
    when exploration is performed on \textit{commodity hardware}.

    \item \textbf{Approximate Analysis.}
      In numerous exploratory and analytical tasks, users often prioritize speed over exact results.
      Developing efficient approximate mechanisms that operate (on-the-fly) on large raw files while offering \textit{adjustable accuracy and ensuring reliable error guarantees}, is highly challenging.

\end{itemize}

\stitle{Our Approach.}
To address these challenges, we propose an \textit{approximate query processing framework} that balances accuracy and efficiency for large raw data files.
In this context, the framework incorporates techniques such as \textit{on-the-fly index construction}, integration of \textit{adaptive indexing with incremental user-driven sampling}, and \textit{incremental aggregate computations}.

\begin{tcolorbox}
	[colback=gray!30,colframe=white,arc=0pt,outer arc=0pt,
left=4pt, right=4pt, top=5pt, bottom=5pt]
The proposed framework  ensures \textit{low data-to-analysis time}, \textit{high performance} even in challenges cases (e.g., high object density), and \textit{efficient approximate computations}.
The effectiveness of the proposed framework against the aforementioned challenges is clearly  demonstrated in the experimental analysis section.
\end{tcolorbox}

\stitle{Contributions.}
In this work, we propose an \textit{adaptive approximate query processing framework} for raw data exploration that combines \textit{incremental sampling} with \textit{adaptive indexing} to balance perfomance  and accuracy. Our main contributions are:

\begin{itemize}

\item
We introduce a  main-memory \textit{adaptive indexing scheme} (\ind) that is designed for efficient approximate query processing.

\item We introduce a \textit{user-driven incremental sampling} approach that is integrated with  index adaptation techniques.

\item We implement  a mechanism that \textit{incrementally compute and reuse partial aggregates}
derived from sampled data, reducing redundant file accesses and improving efficiency.

\item We introduce  an \textit{error-bounded approximation strategy}
that maintains confidence guarantees while reducing I/O operations.

\item We implement and evaluate our approach in an {in-situ} query processing environment,
demonstrating significant improvements in query latency and I/O efficiency, using real and synthetic datasets.

\end{itemize}

\stitle{Outline.}
The paper is organized as follows.
Section~\ref{sec:background} introduces the fundamental concepts of our framework, including the exploration model and the indexing approach we build upon.
Section~\ref{sec:index} formally describes \ind for approximate query processing, while Section~\ref{sec:aqp} details the query execution workflow, integrating approximate query answering and index adaptation.
Section~\ref{sec:eval} presents the experimental evaluation, and Section~\ref{sec:rw} reviews related work.
Finally, Section~\ref{sec:concl} concludes the paper.


\section{Background}
\label{sec:background}

This section outlines the exploration model that this work based on and provides a brief overview of the tile-based \valinor index \cite{IS}, which serves as the foundation for our work.

\subsection{Exploration Model}
\label{sec:model}

\stitle{Data File \& Data Objects.}
We consider the scenario where a user interactively explores data stored in a large \textit{raw data file} \( \F \) (e.g., CSV) using 2D visualization techniques, such as scatter plots or maps. 
The raw data file  consists of a set of \( d \)-\textit{dimensional} \textit{objects} \( \O \), where each object \( o_i \) is represented as a list of attribute values:
$o_i = (a_{i,1}, a_{i,2}, \dots, a_{i,d})$.
Each attribute \( A \in \A \) may be spatial, numeric, or textual.
Furthermore, each object \( o_i \) is  associated with an \textit{offset} \( f_i \), a hex value, indicating its position within the file $\F$.

The dataset includes at least two numeric attributes (e.g., longitude, latitude), which are mapped to the X and Y axes of the visualization and are referred to as \textit{axis attributes} $A_x$ and $A_y$. The remaining attributes are referred as  \textit{non-axis attributes}.

\begin{myExample}
\textbf{[Data Objects]}
\label{ex:object}
In Figure~\ref{fig:index}a sample of the raw data is presented,
containing five objects ($o_1-o_5$), where each object represents a sky object, such as a star (as described Sec.~\ref{sec:intro}).
For example, considering the object $o_1$, we have that $a_{1,1} = 21$, $a_{1,4} = 7$, etc.
Further, for each object $o_i$ there is a file pointer $f_i$ that corresponds to the \textit{offset of $o_i$ from the beginning of the file}.
\end{myExample}


\begin{table}[H]
\centering
\caption{Common Notation}
\label{tab:notation}
\footnotesize
 \SetTblrInner{rowsep=2pt,colsep=4pt}
\begin{tblr}{
   column {1} = {c},
  column {2} = {l},
  row{1} = {abovesep=4pt, belowsep=4pt, font=\bfseries},
  row{even}={bg=gray!15}
}

\hline[0.8pt]

\textbf{Symbol} & \textbf{Description} \\

\hline[0.5pt]
\hline[0.5pt]
$\O$, $o_i$ & Set of objects, an object \\

$\A$, $a_{i,A}$  &  Set of attributes, the value of attribute $A$ of the object $o_i$ \\

 $A_x$, $A_y$ & $X$ $Y$ Axis attributes \\

$Q$, $\R$ & Exploratory query, its Results\\

$Q.I_X$, $Q.I_Y$ & Intervals of query $Q$\\

$\epsilon_{\text{max}}$ &  User-defined error bound \\

$\gamma$  & Query confidence level \\

\( f(A) \)  & Aggregate function applied to attribute \( A \) \\

$\O_Q$ & Objects resulted by the query $Q$\\

$t.I_x$,  $t.I_y$ & Intervals of tile $t$ \\

$t.\O$ &  Objects in tile $t$ \\

$t.\M$   &   Metadata  of tile $t$ \\

$t.\pmb{b}$  &   Bitmap of tile $t$\\

$CI_{\gamma}( f_{i}(A_i))$ &  Confidence interval with probability~$\gamma$   \\

\( \hat{v}_{ f_{i}(A_i) } \) &  Estimated aggregate value for \( f_i \) over \( A_{i} \) \\

\hline[0.8pt]

\end{tblr}
\end{table}

\stitle{User Interactions.} In our exploration model, users interact with the dataset through a set of \textit{user operations} (e.g.,~zoom), which are mapped to data-access operations. These operations define the exploration process, allowing users to dynamically refine their view.

A visualized area \( \Phi = (I_x, I_y) \) represents the current viewport, defined by two numeric intervals $I_x = [x_1, x_2]$ and $I_y = [y_1, y_2]$,
where \( I_x \) and \( I_y \) specify the range of values for the axis attributes \( A_x \) and \( A_y \), respectively.
The visible objects in this area  are those whose axis attribute values fall within these intervals. The mapping between data values and the visualization plane follows a linear or affine transformation.

Users explore the data using operations such as \textit{panning}, which moves the visualized area by shifting \( I_x \) and \( I_y \), i.e., changing the viewport, and \textit{zooming}, which expands or contracts \( I_x \) and \( I_y \) around a focal point, adjusting the level of detail.
As users \textit{explore iteratively}, these operations are often executed in sequence, forming a user exploration session, where each new interaction refines the previous results.
Furthermore, in this work, we also focus on a scenario where the user seeks to analyze the data by computing aggregate values (e.g., mean, sum) over non-axis attributes for the objects in the visualized area.

\stitle{I/O Operation.}
An I/O operation is a file access that  reads the attributes values (even a subset of its attributes) of an object; i.e., the number of I/O operations corresponds to the number of objects, we read their attribute values.

\subsection{VALINOR Indexing Scheme}
\label{sec:valinor}
The locality-based nature of 2D exploratory analysis has been a key consideration in our past work \cite{IS} on in-situ visual exploration, where we introduced \valinor.
It is a \textit{hierarchical, tile-based indexing scheme} designed to optimize query performance by progressively adapting to user interactions. The goal of \valinor is to efficiently support dynamic, interactive exploration and statistics computations over large raw data files while minimizing I/O costs.

\valinor is an in-memory index that organizes data objects into hierarchies of non-overlapping rectangular tiles. These tiles are defined over the domains of the axis attributes \( A_x \) and \( A_y \). Each tile stores the objects that fall within its boundaries based on their axis attribute values and is associated with a set of metadata (e.g., count, sum, average). Metadata are computed from all objects in a tile and facilitate retrieval of \textit{exact} aggregate values over non-axis attributes.
 The index is initialized and progressively adjusts itself to the user interactions, by splitting the tiles visited  into more fine-grained ones, thus forming a hierarchy of tiles.

During the \textit{initialization phase}, a lightweight, coarse-grained version of the \valinor is constructed on-the-fly, by parsing the raw file once. This results in a small overhead in the  \textit{data-to-analysis time}.

Then, during \textit{exploration}, \valinor is progressively refined based on user interactions.
This is achieved through tile splitting, where tiles that are frequently accessed are dynamically subdivided into finer-grained subtiles enriched with aggregated metadata computed from the file. This progressive index adaptation aims to reduce the number of required I/O operations by aligning the index structure with user exploration patterns.

Due to the exact computation of metadata in each tile, some of the shortcomings of VALINOR are that I/O costs remain high over regions with a high density of objects or during the initial stages of a user exploration session when the index has not yet sufficiently adapted to user exploration patterns (i.e., first queries).
The performance in the aforementioned cases become even more challenging, when the user requests aggregates about attributes whose metadata are not yet stored in the index, e.g., attributes that were not queried in previous interactions.


 \section{\ind: Adaptive Index for   Approximate Query Processing}
 \label{sec:index}

 In this section, we introduce the \ind (\textit{Adaptive  Index for Approximate Query Processing}) indexing scheme. The proposed index extends \valinor with new capabilities for approximate query processing, incremental sampling and an adaptation mechanism (more details in Sec.~\ref{sec:aqp}) which result to a \textit{user-driven sampling strategy}.
 \ind enables  efficient approximate estimation of aggregate statistics   while controlling error bounds.
 Next, we provide some definitions and present the structure and basic concepts of the index.

\subsection{Exploratory Query and Results}
\label{sec:index_preliminaries}

\stitle{Exploratory Query.}
An \textit{exploratory query} \( Q \) over a set of objects is defined by: a \textit{selection area}; a set of \textit{aggregate functions}; a \textit{user-defined error bound}; and a \textit{confidence level}.\footnote{For simplicity, this work considers a simplified version of exploratory queries compared to \cite{IS,MaroulisS22f}, omitting filters and not allowing users to request additional objects details.}
Formally, \textit{exploratory query} \( Q \) is  defined  as tuple:

\mbox{$Q = \langle I_x, I_y, \mathcal{L}, \epsilon_{\text{max}}, \gamma \rangle$}, where

\begin{itemize}

    \item \( I_x = [x_1, x_2] \) and \( I_y = [y_1, y_2] \) define the \textit{query area} (i.e., 2D window), specifying a rectangular range over the axis attributes \( A_x \) and \( A_y \). The query result includes the objects
    \( \O_Q  \subseteq \O \) whose values for \( A_x \) and \( A_y \) fall within these intervals.
    For a query $Q$, we refer to its internals $I_x$ and $I_y$ as $Q.I_x$ and $Q.I_y$, respectively.

    \item  $\L$ is a set $\L= \{ f_i (A_i) \mid 1 \leq i \leq k \}$, where $f_i (A_i)$  denote an algebraic aggregate function $f_{i}$ (e.g., sum, mean) \cite{GrayCBLRVPP97} applied to a numeric non-axis attribute \( A_i \) over the $\O_Q$ objects.
    Note that, in this work we consider only univariate aggregate functions.

    \item \( \epsilon_{\text{max}} \in [0,1)\) is a \textit{user-defined error bound}, specifying the maximum allowable relative error for the approximate result.
Note that, if $\epsilon_{\text{max}} = 0$ (i.e., exact query evaluation),
the index behaves as the original VALINOR index \cite{IS}.

    \item \( \gamma \) is the \textit{confidence level}, ensuring that the actual aggregate value of an aggregate function $f(A_i)$ lies within the computed \textit{confidence interval}
    \( CI_{\gamma}( f(A_i) ) \) \textit{with probability} \( \gamma \).
\end{itemize}

\newpage

\stitle{Query Result.}
The \textit{result} \( \mathcal{R} \) of an exploratory query \( Q \) over \( \O \) is defined as\footnote{For ease of presentation, we define the results to include only the aggregate values, omitting the objects selected by the query.}:

$\mathcal{R} = \{ \langle \hat{v}_{f_i (A_i) }, CI_{\gamma}( f_i (A_i) ) \rangle \mid  \forall f_i (A_i) \in \mathcal{L} \}$,
where

\begin{itemize}

    \item \( \hat{v}_{ f_i (A_i) } \) is the \textit{estimated aggregate} value for \( f_i \) over \( A_{i} \), compute over $\O_Q$ objects.

    \item $CI_{\gamma}\bigl(f_i(A_i)\bigr)$ denotes the \textit{confidence interval}
    \linebreak ${[L_{f_i(A_i)}, U_{f_i(A_i)}]}$ for \(f_i(A_i)\),  such that the exact aggregate value lies within this interval with \textit{probability}~$\gamma$,
    where $L_{f_i(A_i)}$ and $U_{f_i(A_i)}$ are the minimum and the maximum values of $f_i(A_i)$, respectively.

    For exact queries (\(\epsilon_{\text{max}} = 0\)), the confidence interval collapses to a single value, meaning the estimated value is exact. For approximate queries (\(\epsilon_{\text{max}} > 0\)), the confidence interval provides a bounded estimate with controlled uncertainty.

\end{itemize}

\subsection{Index Structure}

\stitle{Tile.}
The index's tiles $\T$ are defined over the domains of the axis $A_x$ and $A_y$ attributes and a \textit{tile} $t \in \T$ is defined from two ranges $t.I_x$ and $t.I_y$, in the same domains, respectively.
Each tile \textit{encloses an ordered set of objects $t.\O$}, when the values
$a_{i,{x}}$  and $a_{i,{y}}$  of an object $o_i \in t.\O$  fall within the
intervals, $t.I_x$ and $t.I_y$ of the tile, respectively.
As $t.\O_{[i]}$ we denote the object in $i$-th position of the objects ordered set.

\stitle{Tiles Hierarchy.}
In each level of the  hierarchy, there are no overlaps between the tile intervals of the same level, i.e., disjoint tiles. A \textit{non-leaf tile} $t$ can have an arbitrary number of \textit{child tiles} $t.\C$, enclosing the intervals of its children. That is, given a non-leaf tile $t$ defined by the intervals ${t.I_x=[x_1,x_2)}$ and  ${t.I_y=[y_1,y_2)}$;
for each child tile $t'$ of $t$, with  $t'.I_x=[x'_1,x'_2)$ and  $t'.I_y=[y'_1,y'_2)$, it
holds that ${x_1 \leq x'_1}$,  ${x_2 \geq x'_2, y_1 \leq y'_1}$ and $y_2 \geq y'_2$.
The \textit{leaf tiles} correspond to tiles without children and can appear at different levels in the hierarchy.

 \begin{myExample}
 \textbf{[Tiles \& Tiles Hierarchy]}
  Considering the input data (Fig.~\ref{fig:index}a),
  Figure~\ref{fig:index}b presents a version of the index,
where the Asc and Decl have been selected as the two axis attributes.
The index (in the upper-level) divides the 2D space into
$4 \times 3$ equally sized disjoint tiles\footnote{For simplicity here we present equally sized tiles.},
and the tile $t_j$ is further divided into $2 \times 2$ subtitles of arbitrary
sizes, forming a hierarchy of tiles.
\end{myExample}

\stitle{Fully-contained and Partially-contained Tile.}
Based on the spatial relationship between the 2D area $Q.I_x \times Q.I_y$ defined by a query $Q$ and an overlapping tile, we classify the tile as either \textit{partially-contained} or \textit{fully-contained} within the region  defined by the query.

Formally, we refer that an interval $I =[a,b]$ is contained into an interval $I'=[c,d]$, denoted as $I \subseteq I'$, when $a \geq c$ and $b \leq d$.
Also, the intersection between two intervals $I$ and  $I'$, denoted as $I \cap I'$,  results to the   interval  $[max(a,c), min(b,d)]$.
An empty interval is denoted as $\varnothing$.

A \textit{tile $t$ is fully-contained in a query $Q$}, when
\mbox{$t.I_x \subseteq Q.I_x$} and  $t.I_y \subseteq Q.I_y$.
Thus, in this case all the tile's objects $t.\O$ contribute to the query result.

Further, a \textit{tile $t$ that is not fully-contained, is said to be partially-contained in a query $Q$}, i.e.,
 $t.I_x \cap Q.I_x \neq \varnothing$ and/or $t.I_y \cap Q.I_y \neq \varnothing$.

\stitle{Tile Metadata.}
Each tile $t$ is associated with a set of aggregate \textit{metadata} $t.\M$ (Fig.~\ref{fig:index}c), consisting of numeric values computed using algebraic aggregate functions over one attribute of the objects in $t.\O$.
These functions include, but are not limited to, \textit{count}, \textit{sum}, \textit{mean}, and \textit{sum of squares of deltas}, enabling efficient computation of aggregates across tiles.

Unlike the exact query answering setting, where metadata store exact aggregate values for all objects in a tile, the \textit{approximate setting maintains metadata for a subset of
 objects}. That is, rather than computing aggregates over all objects in $t.\O$, metadata are computed and updated based only on a \textit{sampled set} $t.\S \subseteq t.\O$, where $t.\S$ consists of objects whose attributes values were read from the file.

To facilitate incremental sampling and metadata updates, each tile $t$ is associated with a \textit{bitmap} $t.\pmb{b}$ of size $|t.\O|$, where each bit indicates whether the corresponding object has been included in the sampled subset $t.\S$.
Specifically:
\begin{itemize}
    \item If $t.\pmb{b}[i] = 1$, the $i$-th object $t.\O_{[i]}$ has been sampled and used in the computation of approximate aggregate values.
    \item Otherwise ($t.\pmb{b}[i] = 0$),  the object $t.\O_{[i]}$  has not been sampled and can be used  in a following sampling  if needed.

\end{itemize}

\stitle{Exact, Approximate  \& Not available Metadata.}
Utilizing the tile bitmap, we can determine whether the metadata  stored in the tile are \textit{exact}, \textit{approximate} or \textit{available}.
Particularly, the \textit{status of a tile $t$ metadata}  can be:
$(1)$~computed over all the objects $t.\O$ included in $t$ (\textit{exact metadata});
$(2)$~computed over a sampled subset of the objects included in $t$ (\textit{approximate metadata}); and
$(3)$~\textit{not available}.
Hence, if all bitmap elements are equal to 1, this indicates that all objects have been read from the file, and the computed aggregates are exact. If all elements are equal to 0, the metadata is not available. If some elements are equal to~1, the metadata is computed approximately.
For example, in Figure~\ref{fig:index}c, based on the bitmap contents,  the objects $o_1$ an $o_5$ have been read from the file. So, the metadata is computed based on these two objects (approximate metadata).

\newpage

\stitle{Tile Contents Summary.}
Figure~\ref{fig:index}c presents the contents of the tile $t_z$.
For each tile $t$, the index stores:
$(1)$ the \textit{intervals} $t.I_{x}$ and $t.I_{y}$;
$(2)$ the \textit{object entries} $t.\O$ contained in the tile, where each
entry contains the values of the axis attributes along with the offset
pointing to the position of the object in the file;
$(3)$ a set of \textit{metadata} $t.\M$ which contains approximate aggregated values incrementally updated over the sampled objects, and  a \textit{bitmap} indicating the tile's object used in aggregate values' computations;
$(4)$ a set of \textit{child tiles} pointers $t_z.\C$.

\stitle{Metadata, Computations \& Sampling.}
In nutshell, the use of metadata during query processing is outlined as follow (more details in Sec.~\ref{sec:query_processing}).
If a tile's  \textit{metadata is exact}, its values are used to estimate the tile's contribution to the result.
Else, if the tile's \textit{metadata is approximate}, the  approximate aggregates are used in computations.
Here, there are cases that the approximate metadata results to a confidence interval that does not meet the \textit{user-defined error threshold}.
In these cases, additional samples are \textit{incrementally drawn from the unsampled objects} and used to refine existing metadata.
This process incrementally refines both the \textit{aggregate estimates and the confidence interval until the error bound is satisfied}.
Finally, in case the tile's \textit{metadata is not available}, the same incremental sampling  procedure as in the previous case is followed.

\subsection{Index Initialization}
The initialization process for
\ind follows the same principles as the original VALINOR index~\cite{IS}. Instead of a separate loading phase, the index is constructed \textit{on-the-fly} when the first query \(Q_0\) is issued, ensuring minimal  \textit{data-to-analysis time} overhead. The raw file is parsed once to create an initial flat tile structure while evaluating the first query. In the simplest version, tile sizes are determined using a binning technique that partitions the data space into equal-sized tiles, forming a lightweight tile-based index.

During parsing, objects are assigned to their respective tiles, and  metadata for non-axis attributes is computed and stored. At the end of initialization, each tile contains \textit{exact metadata}, pertaining to all its objects.

More advanced initialization strategies, such as \textit{query-driven initialization}~\cite{IS}, refine the tile layout by allocating smaller tiles near the first query. This increases the likelihood of fully contained tiles in early exploration steps, reducing file access and improving query performance. For further details, we refer the reader to~\cite{IS}.


\section{Approximate Query Processing \& Index Adaptation}
\label{sec:aqp}

In this section, we present our methodology for the evaluation of an exploratory query over the \ind, the user-driven sampling strategy for computing aggregate statistics and the index adaptation process.
We first provide an overview of our methods, and proceed with the detailed steps of the query processing.

\subsection{Query Processing Process}
\label{sec:query_processing}

Given an exploratory query \( Q = \langle I_x, I_y, \mathcal{L}, \epsilon_{\text{max}}, \gamma \rangle \), our approach evaluates the query by identifying the tiles overlapping the region defined by the query, leveraging precomputed metadata, and employing incremental sampling to estimate aggregates while ensuring error guarantees. The query evaluation consists of the following steps.

Note that the following steps are not performed in a strict order, as some steps occur in parallel during the incremental sampling.
As described next, the Steps 3--5 are executed iteratively, since are parts of the  incremental sampling procedure.
The query processing process is also described  in Example~\ref{ex:qp}.

\subsubsection{[Step 1] Find Overlapping Tiles}
\label{Sec:step1}

We first identify the set of overlapping tiles by determining their spatial relationship with the query region \mbox{\( Q.I_x \times Q.I_y \)}. These tiles are either \textit{fully-contained} or \textit{partially-contained} by the query.

\subsubsection{\textnormal{\textbf{[Step 2]}} Combine Spatial Relations and Metadata Status to Access Required Sampling}
\label{Sec:step2}

For each tile overlapping the query region, we assess its \textit{spatial relation to the query} and \textit{its stored metadata status}, to decide if additional objects must be retrieved (sampled) from the file.
We have the following \textit{four  cases}:

\begin{itemize}
    \item \textbf{Case 1:} \textit{Fully-contained tile with exact metadata}.
For each fully contained tile, if aggregate metadata exist that represent all objects in the tile, our approach  utilizes them directly, i.e., \textit{eliminating the need to access the file}.
These complete metadata contribute to the query result without introducing any uncertainty.

\item
\textbf{Case 2:} \textit{Fully-contained tile with approximate metadata}.
In this case metadata of a fully-contained tile have been computed on a sampled subset of the tile’s objects. The computed approximate values are used to estimate the query result without the need to read further samples from the file.

However, as analyzed in the next sections, there are cases where \textit{additional sampling is needed}, since the estimated uncertainty exceeds \( \epsilon_{\text{max}} \) (more details in next section).
As a result, file access may be required in order to read the new objects that will be added in the sample. Also in this case, the metadata are incrementally updated based on the new samples.

\item
\textbf{Case 3:} \textit{Partially-contained tile}.
Regardless of the metadata status (whether exact or approximate), metadata cannot be used for partially contained tiles. Therefore,  we \textit{perform sampling in this tile}.

A tile’s metadata pertains to a (subset of) all objects within the tile. However, in this case, we require metadata specifically for the objects that fall within the query region. Thus, sampling is conducted over these objects using an initial sampling rate. This rate is incrementally adjusted until the user-defined error threshold is met (more details follow).
 As a result, \textit{file access is required} only for the sampled objects.

\item
\textbf{Case 4:} \textit{Fully or Partially-contained Tile with no metadata}.
During index initialization or adaptation, new tiles are created. In the particular case of index adaptation (more details in Sec.~\ref{sec:adapt}), a tile overlapping the user query can split into new \textit{fully-contained and/or partially-contained tiles}, with empty metadata. We perform sampling to compute metadata of these new tiles.
Note that, index adaptation is not separate step in query evaluation, but rather it is performed in parallel with the current step.

\end{itemize}

\stitle{User-driven Sampling.}
In our approach, sampling is tightly integrated with index adaptation, forming a \textit{user-driven sampling} approach. Tile splitting increases index granularity in frequently explored regions, reducing the query area that requires sampling.
By increasing the number of fully-contained tiles, our method minimizes file access since such tiles can be answered directly using stored metadata. Furthermore, based on the user-defined error threshold, we retrieve and store metadata only for a subset of objects per tile; yielding partial metadata that is sufficient for answering subsequent queries with similar error bounds without additional I/O.

\subsubsection{ \textnormal{\textbf{[Step 3]}} Read Samples from the File}
\label{Sec:step3}
In this step, we perform sampling on the objects within the tiles identified in the previous step to compute their metadata. Sampling is conducted incrementally, with a sampling rate estimated at each iteration. Since multiple sampling iterations may be required, heuristics are applied to adjust the sampling rate dynamically in each round (more details follow).

To reduce random I/O's and improve file access performance, we adopt the following approach:
we sort objects in the tile based on their offsets before accessing the file, ensuring that the I/O operations are performed in sequential order. This reduces the overhead of fragmented disk I/O operations and enhances overall query efficiency.

\subsubsection{\textnormal{\textbf{[Step 4]}} Metadata Incremental Updates}
\label{Sec:step4}

Each time new objects are read from the file, their values incrementally update the available metadata, ensuring that future queries benefit from more refined estimates. While the updated metadata can be leveraged in the current query to compute a value estimation and a confidence interval, there are cases where it may not remain valid for future queries.

In fully contained tiles, the updated metadata are computed from uniformly sampled objects in the entire tile. Therefore, it remains valid for any query and stored in the index. However, in partially contained tiles, the sampled objects are drawn only from the region overlapping the query, meaning they cannot be used to update the tile's metadata. Although these objects belong to the tile, incorporating them would not ensure a uniform sample across the entire tile. Consequently, while the metadata is used for query computation in this case, it is not stored in the index.

 \subsubsection{\textnormal{\textbf{[Step 5]}} Compute Aggregate Value Estimation \& Confidence Interval}
 \label{sec:step5}
In this step, we combine exact and approximate (sample-based) metadata of tiles to compute a total \textit{aggregate estimate} and its \textit{confidence interval}.

For each aggregate function \( f(A) \in \mathcal{L} \), the \textit{confidence interval}
\( CI_{\gamma}(f(A)) \) is computed by combining exact and approximate metadata from the tiles overlapping the query.
We treat each tile as a \emph{stratum} in a stratified sampling-based approach.

The final estimate follows stratified sampling principles, where the total aggregates consist of:

\begin{itemize}
    \item \textit{Exact metadata} from fully-contained tiles with metadata computed
      over \emph{all} objects in the tile (\textit{Step~2:~Case~1}).
\item \textit{Approximate metadata} from fully-contained tiles whose metadata are
      based on a subset of objects in the tile (\textit{Step~2: Case 2}).
\item \textit{Approximate metadata} from partially contained tiles, where new
      samples are drawn specifically from the portion of each tile overlapping
      the query region (\textit{Step~2: Case 3}).
\end{itemize}

For the sampled portions, we apply the \emph{Central Limit Theorem (CLT)} to approximate the distribution of the mean estimator as normal when the sample size is sufficiently large. The CLT states that, as the sample size increases, the distribution of the sample mean approaches a normal distribution centered at the true population mean, regardless of the underlying data distribution:
\[
\hat{v}_{f(A)} \sim \mathcal{N} \Bigl( \mu, \frac{\sigma^2}{N} \Bigr),
\]

\noindent
where \(\mu\) is the true population mean for the queried region, \(\sigma^2\) is the population variance, and \(N\) is the total number of sampled objects.

\stitle{Aggregate Value Estimation:  Combining  Exact \& \linebreak Approximate Metadata}
Let \( \T_{F_e} \) be the \textit{set of fully contained tiles with exact metadata}, and
\( \T_{F_a} \) be the set of \textit{fully contained tiles with approximate metadata}.
Also,  \( \T_P \) denote the set of \textit{partially contained tiles}, from which new
samples are drawn from their intersection with the query.

\begin{itemize}
    \item
\textbf{Linear, decomposable aggregates} (e.g., sum): the \textit{overall estimated aggregate value} is:
\[
\hat{v}_{\mathrm{sum(A)}}
\;=\;
\sum_{\forall t \in \T_{F_e}} v_t
\;+\;
\sum_{\forall t \in \T_{F_a}} \hat{v}_t
\;+\;
\sum_{\forall t \in \T_P} \hat{v}_t,
\]
where \( v_t \) is the exact aggregate value for a fully contained tile \( t \in \T_{F_e} \),
and \( \hat{v}_t \) is the approximate aggregate for a tile
\( t \in \T_{F_a} \cup \T_P \), computed via sample-based estimation.
Because sums are \emph{additive}, we can simply add the tile-level results.

\item
\textbf{Mean  aggregate:} each fully or partially contained tile provides a
\emph{partial sum} (exact or approximate, based on the metadata status)
and an \emph{object count} for the query region. To obtain the overall
mean, we add these partial sums from all tiles and then divide by the total
object count across them.

\item
\textbf{Count  aggregate:} as we store the $A_x$ and $A_y$ attribute values
of each object in our index, we can  directly identify all objects contained in the query region.
Consequently, \({count}\)  does not require an approximation.

\end{itemize}

\stitle{Variance Estimation.}
Since our approach follows a stratified sampling model (treating each tile as a
stratum) and we sample \emph{without replacement},   let:

\begin{itemize}
  \item \(N_t\) = \emph{population size} of tile~\(t\), i.e., number of objects
        in the tile $t$ that lie in the queried region
  \item \(n_t\) = \emph{sample size} in tile~\(t\) (how many objects we actually read)
  \item \(\hat{\sigma}_t^2\) = \emph{sample variance} of the attribute in tile~\(t\)
  \item \(N = \sum_{\forall t \in \T_Q} N_t\) = \textit{total population} in the query region,
  where $\T_Q$ are the tiles overlapped with the query.

\end{itemize}

\noindent
For a \emph{{\textbf{sum}}} over the queried region, each tile’s estimated contribution is
\(\hat{v}_t \!= N_t\,\hat{\mu}_t\), where \(\hat{\mu}_t\) is the sample mean of tile~\(t\).
A standard stratified‐sampling formula with the Finite‐Population Correction (FPC) is:
\[
  \mathrm{Var}\bigl(\hat{V}_{\mathrm{sum}}\bigr)
  \;=\;
 \underset{\substack{\forall t  \in \\ (\T_{F_a}\cup \T_P)}}{\sum}
    N_t^2 \,\frac{\hat{\sigma}_t^2}{\,n_t\,}
    \;\Bigl(1 - \tfrac{n_t}{N_t}\Bigr)
\]

\noindent
For a \emph{\textbf{mean}} over the region, the overall estimator is \linebreak
\({\hat{V_\mu} \,= \sum_t \bigl(\tfrac{N_t}{N}\bigr)\,\hat{\mu}_t}\).
Hence,
\[
  \mathrm{Var}\bigl(\hat{V_\mu}\bigr)
  \;=\;
 \underset{\substack{\forall t  \in \\ (\T_{F_a}\cup \T_P)}}{\sum}
    \Bigl(\tfrac{N_t}{N}\Bigr)^{2}\,
    \frac{\hat{\sigma}_t^2}{\,n_t\,}
    \;\Bigl(1 - \tfrac{n_t}{N_t}\Bigr)
\]

Note that, we include FPC in the above formulas because we sample \emph{without replacement}.
This choice aligns with our \textit{incremental sampling approach}: as we read more objects from a tile to refine its metadata, we may eventually exhaust the tile, obtaining \emph{exact metadata} and thus eliminating uncertainty for future queries on that tile.

\stitle{Confidence Interval Computation.}
Using the CLT, we then derive a \textit{confidence interval} for either \textit{\textbf{sum}} or \textit{\textbf{mean}} in the standard way, for \textit{confidence level}~\(\gamma\):
\[
  CI_{\gamma}(f(A))
  \;=\;
  \Bigl[
    \hat{v}_{f(A)}
    \;\pm\;
    z_{\gamma}\,\sqrt{\mathrm{Var}(\hat{v}_{f(A)})}
  \, \Bigr],
\]
where \(z_\gamma\) is the  appropriate normal quantile.

Note that, because exact contributions from \(\T_{F_e}\) have zero variance, they do not
increase uncertainty in the final estimate.

In the case of \emph{\textbf{variance}} or \emph{\textbf{standard deviation}} aggregate function, additional
adjustments via the chi-square distribution are required (omitted for brevity).

Regarding, \emph{\textbf{min}} and \emph{\textbf{max}}, confidence intervals are generally unreliable
because sampling may miss extreme values; so, in those cases, the index falls
back to exact computation.

  \begin{figure*}[t]
 	\centering
\vspace{6pt}
\includegraphics[width=0.95\linewidth]{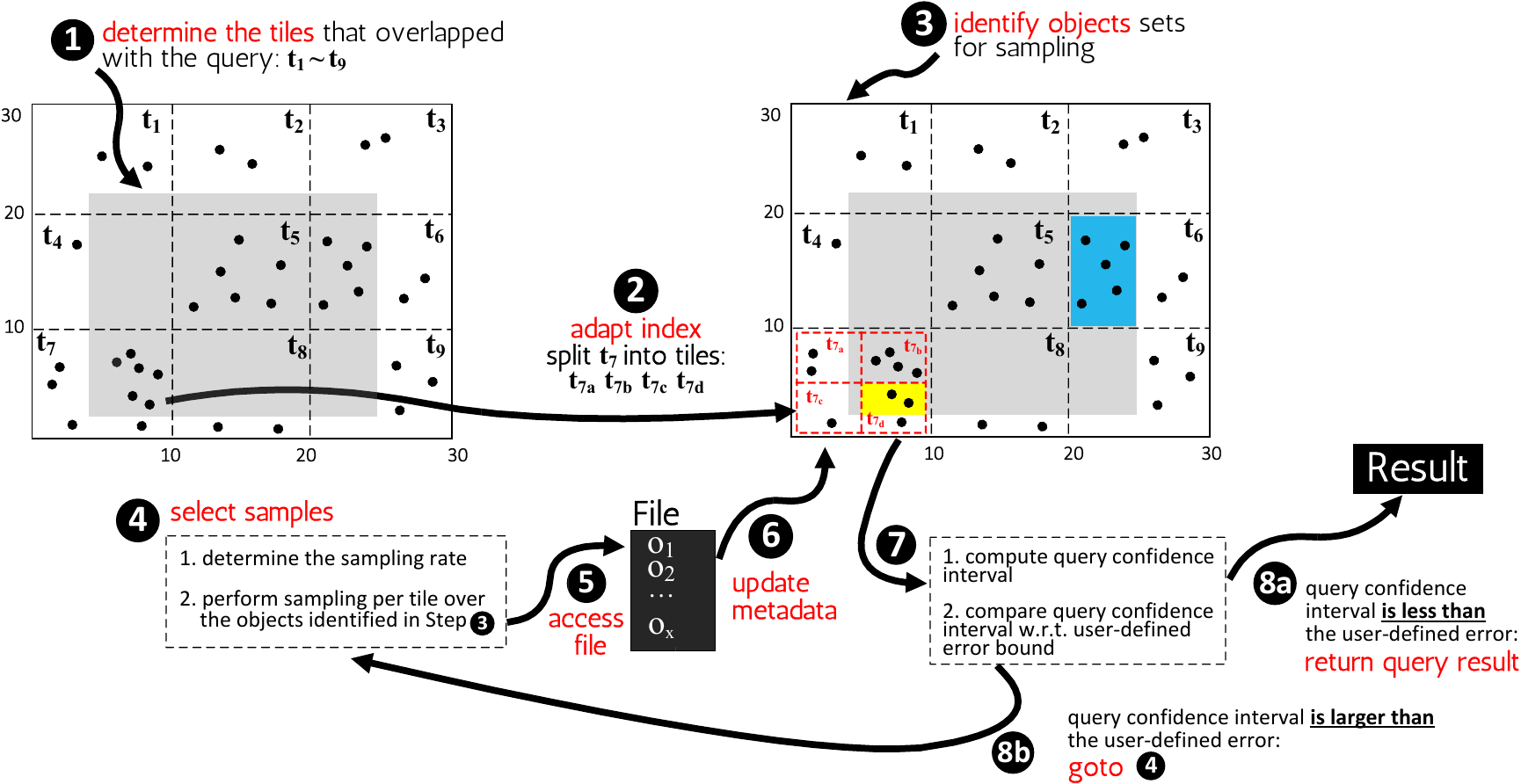 }
 \caption{Approximate Query Processing \& Index Adaptation}
 \label{fig:query_proc}
  \end{figure*}

\subsubsection{Incremental Sampling Procedure}
\label{sec:incrsample}

In our incremental sampling process, each iteration involves  the Steps 3--5.
Particularly, after deriving the confidence interval (Step~5), our  approach checks whether the
\emph{maximum relative error} \(\epsilon_{\text{est}}\) satisfies the \textit{user-defined bound}
\(\epsilon_{\text{max}}\).

If the user-defined bound is not satisfied, a next sampling iteration is performed.
So, the execution goes to Step 3, estimating a new sampling rate as described next.
The process continues incrementally, refining query estimates until the computed error falls within the acceptable range defined by \( \epsilon_{\text{max}} \).
Otherwise, the query result is formed and returned to the user.

\sstitle{\underline{Relative Error}.}
The \emph{maximum relative error} is computed by taking
\emph{half the confidence interval's width} (the margin of error)
over the estimated value (the midpoint of the interval):
\[
  \epsilon_{\text{est}}
  \;=\;
  \frac{
    \tfrac12 \,\bigl\lvert CI_{\gamma}(f(A)) \bigr\rvert
  }{
    \hat{v}_{f(A)}
  }
\]
Here, \(\hat{v}_{f(A) }\) is the estimated aggregate value, and
\(\bigl\lvert CI_{\gamma}(f(A)) \bigr\rvert\) denotes the interval's total width.
If \(\epsilon_{\text{est}} > \epsilon_{\text{max}}\), additional sampling is triggered
to reduce uncertainty, typically by reading more objects from tiles where
metadata remain incomplete. The process continues until
\(\epsilon_{\text{est}} \le \epsilon_{\text{max}}\).

\stitle{Sampling Rate Adjustment.}
To reduce error efficiently, our approach \emph{adaptively adjusts the sampling rate} rather than using a fixed-step increase. We utilize a \textit{heuristic approach} where, in each iteration, we compute a \emph{multiplicative factor} based on the ratio of the current error to the desired threshold, typically \(\bigl(\tfrac{\epsilon_{\text{current}}}{\epsilon_{\text{max}}}\bigr)^2\), reflecting the \(\tfrac{1}{\sqrt{n}}\) relationship between sample size and standard error. If this factor exceeds a cap (e.g., 2.0), we limit the increase to avoid excessively large jumps in the sampling rate. Conversely, if the computed increment is too small, we enforce a minimum increase to ensure noticeable progress.

This adaptive adjustment balances error reduction with I/O costs. In each sampling iteration, the tiles are sampled based on the current sampling rate, and the objects selected are potentially scattered throughout the file, incurring random I/O overhead. By making larger sampling rate jumps when error is high (and refining the rate more gradually near the threshold), we \emph{reduce the number of rounds} requiring additional data retrieval, thus minimizing repeated random accesses. This strategy ensures that each batch of additional samples contributes significantly to narrowing the confidence interval while avoiding unnecessary data retrieval.

\subsection{Index Adaptation}
\label{sec:adapt}
We employ the index adaptation technique from \cite{IS} that adjusts the index based on user queries and aims at minimizing I/O operations and computational costs, progressively increasing interactivity in frequently explored areas. The index adaptation is performed after a user operation, affects the tiles of the index that are contained in the user query and results in: \textit{modifying the index structure}, i.e., splitting tiles into multiple ones and adjusting tile sizes, and \textit{enriching tiles with missing metadata}.

Adaptation is performed in parallel with Step 2 (Combine Spatial Relations and Metadata Status to Initialize Sampling) (Sec.~\ref{Sec:step2}), where sampling is guided by tiles splitting, resulting to a \textit{user-driven sampling process}.

The  decision to split a tile is guided by I/O cost considerations, ensuring that adaptation remains beneficial for future queries. For each tile we need to split, we estimate the expected splitting gain in
terms of I/O cost, for evaluating a (future) query. If the expected splitting gain exceeds a fixed threshold, a split is performed. A further
analysis of the splitting model is presented in \cite{IS}.
Given the locality-based characteristics of interactive exploration, tile splitting increases the likelihood that subsequent queries will fully contain a tile within a frequently explored region.

In our experiments, we follow a quadtree-like splitting strategy that recursively divides the tile into four equal subregions, for which new statistics are calculated.

\begin{myExample}
\textbf{[Approximate Query Processing \& Index Adaptation]}
\label{ex:qp}
 The query processing and index adaptation process are presented in Figure~\ref{fig:query_proc}.
As input we have an initialized index  where each tile has exact metadata, an exploratory query and a raw file.
 Note that, the steps numbering here differs from the steps numbering in Section~\ref{sec:query_processing}

\circledFill{1}
To evaluate query $Q$  we first  	 \textit{find the leaf
tiles that spatially overlap} (partially or fully-contained)
with the query region:
$t_1$,  $t_2$, $t_3$,... $t_9$. Since no objects are selected by the query from the tiles $t_1$,  $t_2$, $t_{3}$, $t_{4}$, $t_{8}$ and $t_{9}$, these tiles are not included in the query evaluation process.
\circledFill{2}
Next, we \textit{check if the overlapping tiles need to be split},
in such case, the tiles are split into smaller subtiles.
In each splitting step, the process considers criteria related to I/O cost in order to decide
whether to perform a split or not (more details at Sec.~\ref{sec:adapt}).
For simplicity, in our example,  we assume that $t_7$ is split into four equal disjoint subtiles:
$t_{7_a}$, $t_{7_b}$, $t_{7_c}$, and $t_{7_d}$.
Since no objects are selected from the tiles $t_{7_a}$ and $t_{7_c}$, these tiles are not further considered.

\circledFill{3}
Next, we examine each of the overlapped tiles (including  the ones resulted by splitting), to determine which case each tile belongs to (Case 1--5, Sec.~\ref{Sec:step2}).
We have the following tile categorization:
(1) \textit{Case~1 (Fully-contained tile with exact metadata)}: $\mathbf{t_{5}}$;
(2) \textit{Case~3 (Partially-contained tile}): $\mathbf{t_{6}}$; and
(3) \textit{Case~4 (Fully or Partially-contained  tile resulted by  Adaptation)}:
$\mathbf{t_{7_b}}$ and $\mathbf{t_{7_d}}$.


Based on this categorization, for each tile, we determine the  set of objects in which a sampling has to be performed (i.e., read objects attributes' values from the file):
(1)~$\mathbf{t_{5}}$:~we do not need to read objects from the file (no sampling);
(2)~$\mathbf{t_{6}}$:~we have to perform sampling over the objects included in the area selected by the query, i.e., the objects included in blue area;
(3)~$\mathbf{t_{7_b}}$:~we have to perform sampling  considering all the tile's objects; and
(4)~$\mathbf{t_{7_d}}$~we have to perform sampling over the objects included  in the area selected by the query, i.e., the objects included in yellow area.

\circledFill{4}
In tiles $t_{6}$, $t_{7_b}$ and $t_{7_d}$ we use a sampling method to select the objects to read from the file, considering the object sets defined in the previous step.
First, a sampling rate is determined. As described, in our incremental sampling process, a different sampling rate is computed and adjusted in each iteration.
Based on this rate, a uniform sampling is used to select which objects to be read from the file \circledFill{5}.

 \circledFill{6}
Using the data retrieved from the file, the metadata of each tile is computed from scratch or updated. This update incorporates the newly sampled objects, thereby refining the tile’s stored metadata.

 \circledFill{7}
Based on updated metadata,  the query's confidence interval is computed by combining each tile's contribution—using exact metadata when available and approximate metdata otherwise.

The computed query confidence interval is then compared against the user-defined error bound.
\circledFill{\scriptsize{8a}}
If the computed relative error bound is within the acceptable bound, the system returns the query result along with its confidence interval.
\circledFill{\scriptsize {8b}}
Otherwise (the interval remains too wide),
a new  sampling iteration is performed \mbox{Steps 4--7},
and the \mbox{Steps 4--7} are repeated until the error bound is satisfied.

 \end{myExample}

\begin{figure*}[t]
	\centering
	\begin{subfigure}{0.45\textwidth}
		\centering
		\includegraphics[width=\linewidth]{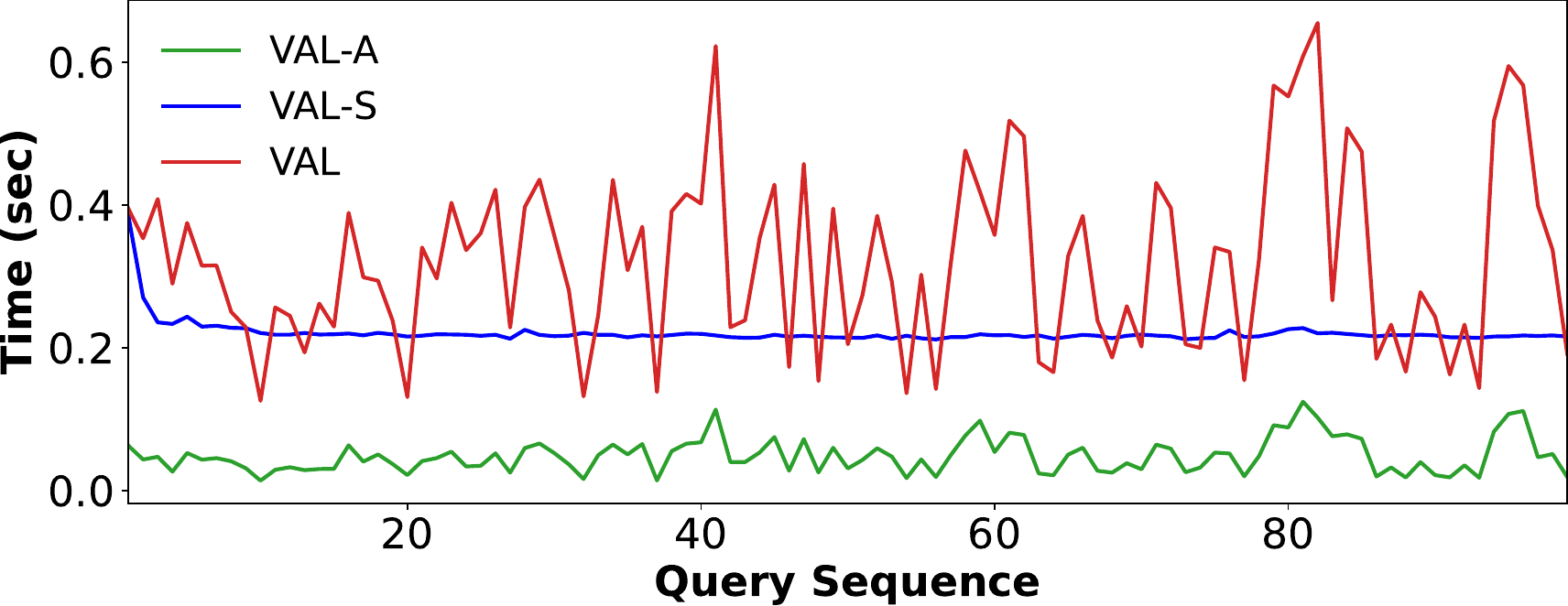}
		\caption{SYNTH10}
		\label{fig:synth10}
	\end{subfigure}
	\hspace{30pt}
	\begin{subfigure}{0.45\textwidth}
		\centering
		\includegraphics[width=\linewidth]{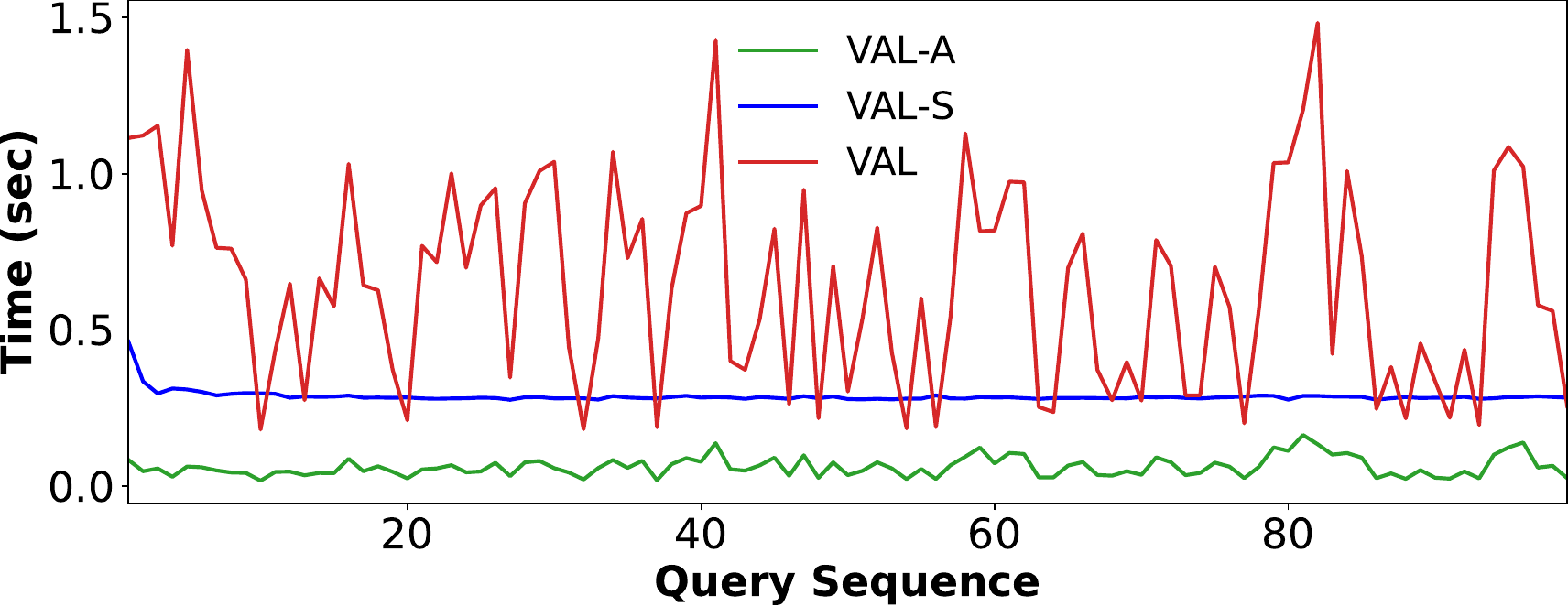}
		\caption{SYNTH50}
		\label{fig:synth50}
	\end{subfigure}

	\vspace{14pt}

	\begin{subfigure}{0.45\textwidth}
		\centering
		\includegraphics[width=\linewidth]{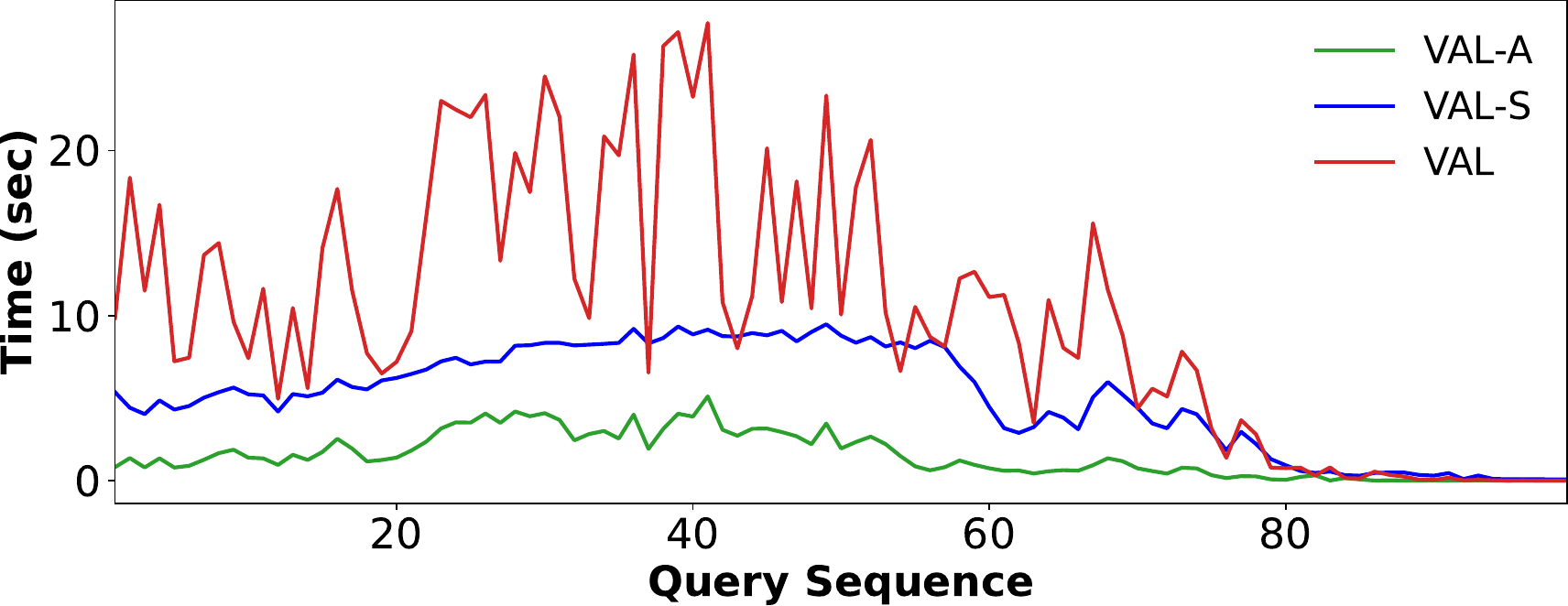}
		\caption{TAXI}
		\label{fig:taxi}
	\end{subfigure}

	\caption{Query execution time per query.}
	\label{pl:workload}
\end{figure*}

\section{Experimental Analysis}
\label{sec:eval}
The objective of our evaluation is to assess the performance and the effectiveness
of our approach in terms of response time, I/O operations, relative error, and confidence interval coverage. We evaluate our index and  competitors over one real and two synthetic datasets.

\subsection{Experimental Setup}
\label{sec:exp_set}

\stitle{Datasets \& Queries.}
 In our experiments we have used two synthetic datasets (SYNTH10 / 50), and one real dataset, the \textit{NYC Yellow Taxi Trip Records} (TAXI).

\begin{itemize}

    \item
 \sstitle{\dunderline[-2pt]{0.5pt}{SYNTH10 \normalfont{/} 50  Synthetic Datasets.}}
Regarding the \textit{synthetic datasets} (SYNTH10 / 50),
we have generated two CSV files of 100M \textit{data objects}, having 10 and 50 \textit{attributes} (11 and 51~GB, respectively).
The datasets contain \textit{numeric attributes} in the range $[0, 1000]$, following a uniform distribution.\footnote{The data generator and the queries are available at: \href{https://github.com/VisualFacts/RawVis}{github.com/VisualFacts/RawVis}}
 Regarding queries, as in \cite{MaroulisS22f,IS}, the query region is defined over two attributes that specify a window size containing  approximately 100K objects.

\item
\sstitle{\dunderline[-2pt]{0.5pt}{TAXI Dataset.}}
The TAXI dataset is a CSV file, containing information regarding taxi rides in NYC.\footnote{ \scalebox{0.95}{\href{https://www1.nyc.gov/site/tlc/about/tlc-trip-record-data.page}{www1.nyc.gov/site/tlc/about/tlc-trip-record-data.page}}}
Each record corresponds to a trip, described by
18 \textit{attributes}. We selected a subset of this dataset for 2014 trips with 165M objects and 26 GB CSV file size.
The  {Longitude} and  {Latitude} of the pick-up location are the \textit{axis attributes} of the exploration. The query region is defined over an area of  \mbox{2km $\times$ 2km} size, with the {first query $Q_0$} posed in central Manhattan.
For aggregate computations, we consider the \textit{total trip amount} as the axis attribute of interest.

\end{itemize}

\stitle{Exploration Scenario.}
In our evaluation, we consider a typical exploration scenario such as the one used in  \cite{MaroulisS22f,IS}.
This scenario attempts to formulate a common user  behavior in 2D
exploration, where the user explores nearby regions using pan and zoom operations
\cite{ZhaoRDWHN18,KalininCZ14,TauheedHSMA12,bcs15,YesilmuratI12,DarFJST96},
such as the ``region-of-interest'' or ``following-a-path'' scenarios, which are commonly used in map-based  exploration.
We generated sequences of 100 overlapping queries, with each region shifted by 10\% (i.e., a pan operation) relative to the previous one in a random direction. Although the shift is random, it is biased toward a high-level trajectory, simulating a "follow-a-path" exploration scenario.
At each step, the user requests aggregate values, such as \emph{sum} or \emph{average}, over one of the non-axis attributes of the dataset within the currently visualized area.

\begin{figure*}[t]
	\centering
	\mbox{
		\begin{subfigure}{0.5\textwidth}
			\centering
			\includegraphics[width=\linewidth]{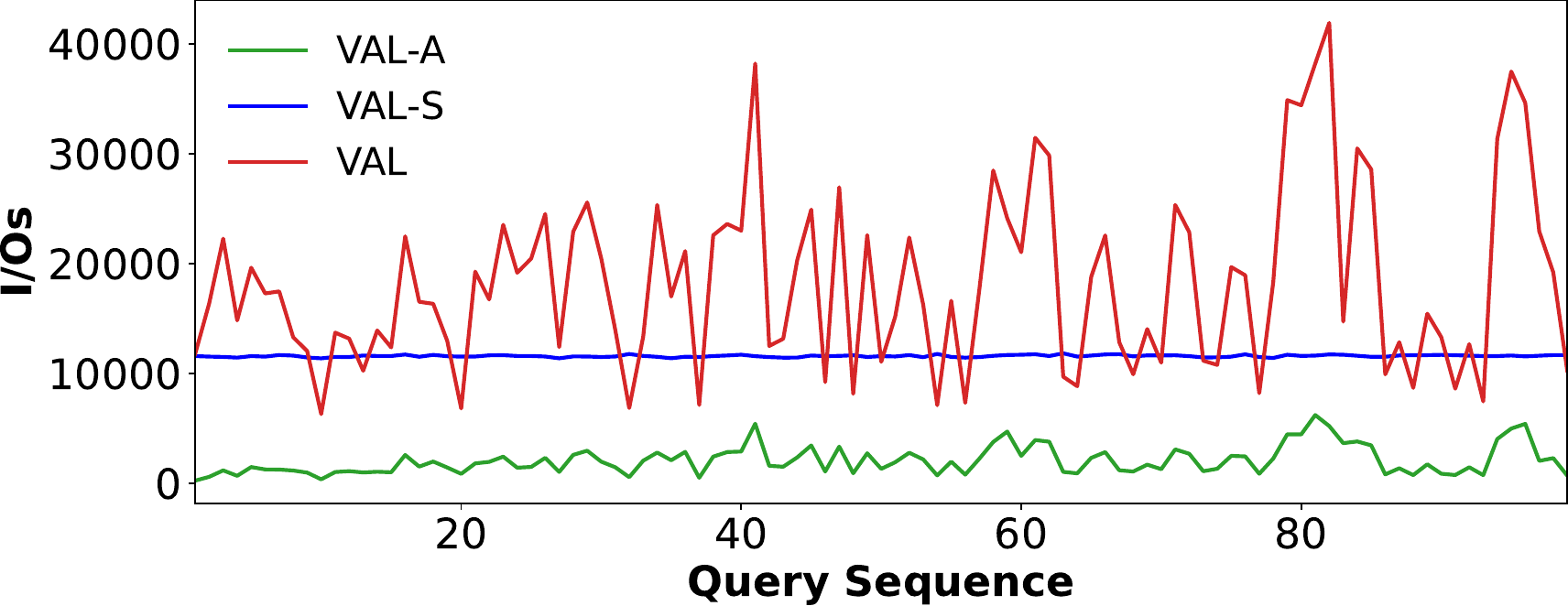}
			\caption{SYNTH10}
			\label{fig:synth10_io}
		\end{subfigure}
		\hfill
		\begin{subfigure}{0.5\textwidth}
			\centering
			\includegraphics[width=\linewidth]{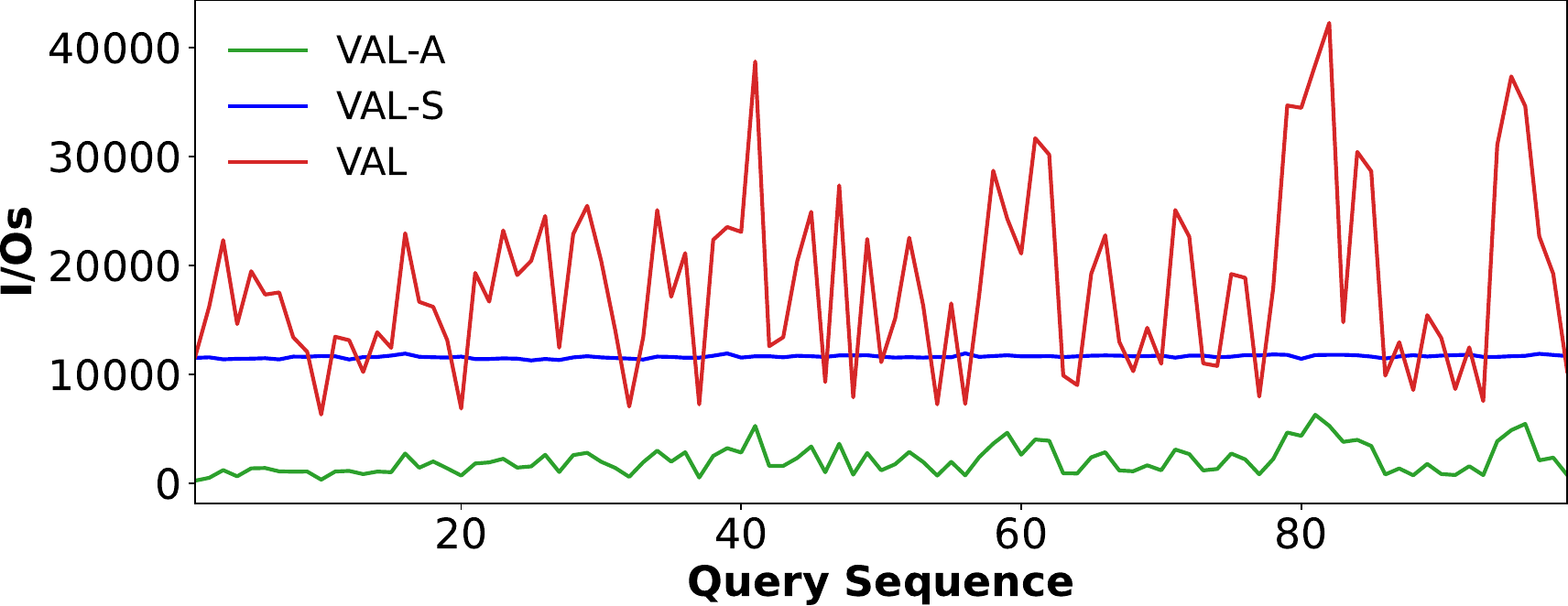}
			\caption{SYNTH50}
			\label{fig:synth50_io}
		\end{subfigure}
	}

	\vspace{14pt}

	\begin{subfigure}{0.5\textwidth}
		\centering
		\includegraphics[width=\linewidth]{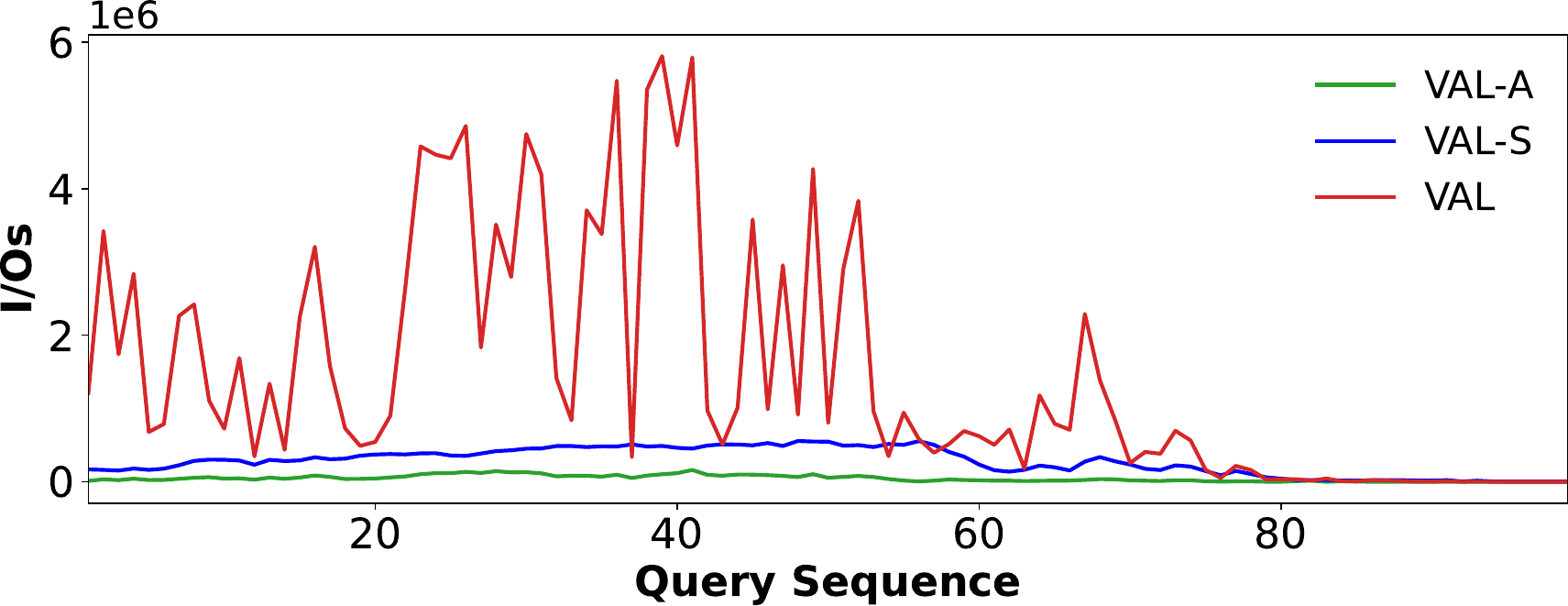}
		\caption{TAXI}
		\label{fig:taxi_io}
	\end{subfigure}

	\caption{Number of I/O operations per query.}
	\label{pl:io}
\end{figure*}

\stitle{Competitors.}
We compare our method  VALINOR-A (\vala) with:
$(1)$~\mbox{VALINOR (\val) \cite{IS}}, which provides exact query answers; and
$(2)$~\mbox{VALINOR-S} (\vals), a baseline that leverages VALINOR indexing to rapidly identify objects within the query region and retrieve them efficiently using stored file offsets. Unlike our approach, \vals does not maintain aggregate metadata; instead, it performs incremental sampling over all objects in the query region for aggregate computations until the error constraint is met. In effect, \vals amounts to a plain, incremental sampling approach without the benefits of reusing precomputed aggregate metadata.

\stitle{Tile Structure Parameters.}
Regarding tile structure, for all the methods, we adopt the setting used in \cite{MaroulisS22f,IS}, where the tile structure is initialized with 100 $\times$ 100 equal-width tiles, while an extra 20\% of the number of initial tiles was also distributed around the first query using the Query-driven initialization method \cite{IS}.
Also, the numeric threshold for the adaptation of \vala was set to 200 objects.
More details about these parameters can be found at \cite{IS}.

\stitle{Metrics.}
In our experiments, we measure  the:
$(1)$~\textit{Execution Time} per query, and \textit{Overall Execution Time} of an exploration scenario, that includes:
initialization time and query evaluation time for all the queries included  in the exploration scenario.
$(2)$~\textit{I/O Operations} performed during query evaluation; and during the whole workflow.
$(3)$~\textit{Relative Error},  defined as the difference between the estimated and exact aggregate values, normalized by the exact value:
\[
\epsilon_{\text{actual}} = \frac{\lvert \hat{v}_{f(A)} - v_{f(A)} \rvert}{v_{f(A)}}
\]
where \( \hat{v}_{f(A)} \) is the estimated aggregate and \( v_{f(A)} \) is the exact aggregate computed from all objects in the query region.
$(4)$~\textit{Confidence Interval Coverage} which measures the proportion of queries where the exact aggregate value falls within the computed confidence interval.

\stitle{Implementation.}
\vala is implemented on JVM as part of the RawVis
\textit{open source data visualization system} \cite{MaroulisBPVV21}.
The experiments were conducted on an 3.60GHz Intel Core i7-3820 with 64GB of RAM.
We applied memory constraints (12GB max Java heap size) in order to measure the performance of our approach and our competitors.

\begin{figure*}[t]
	\centering
	\begin{subfigure}{0.48\textwidth}
		\centering
		\includegraphics[width=\linewidth]{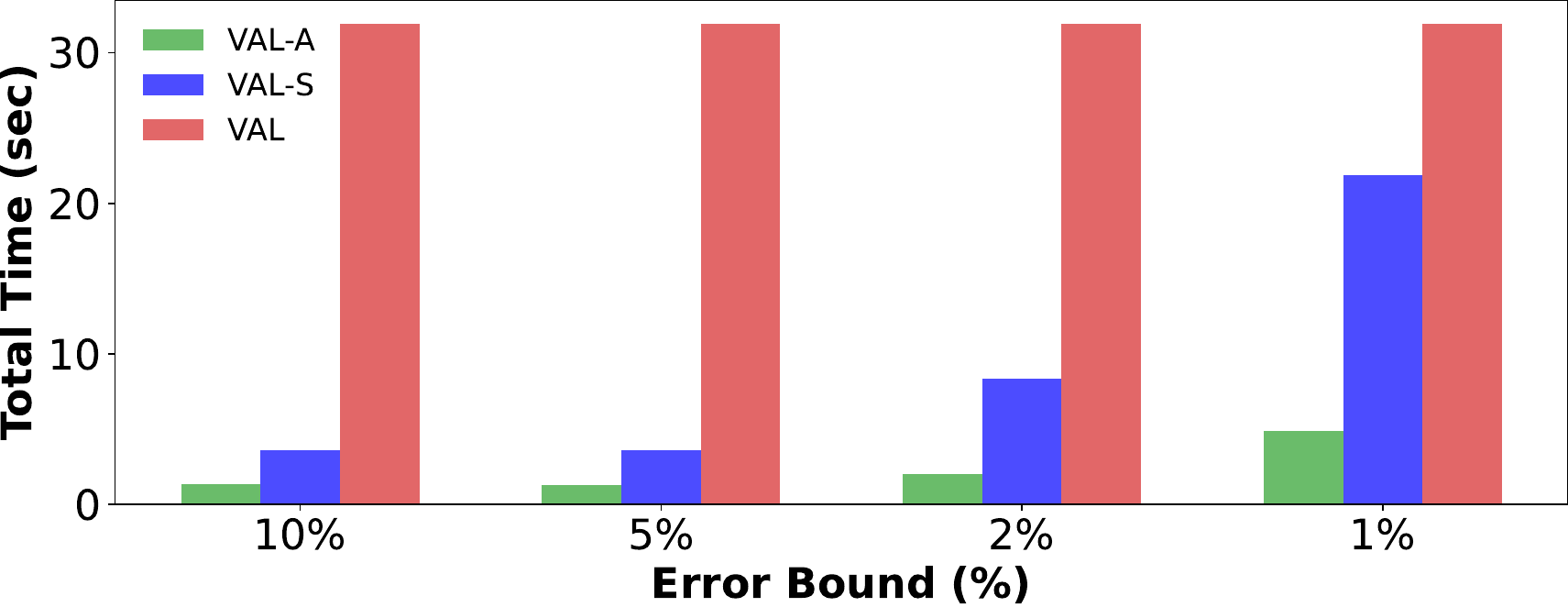}
		\caption{SYNTH10}
		\label{fig:synth10_error}
	\end{subfigure}
	\hspace{10pt}
	\begin{subfigure}{0.48\textwidth}
		\centering
		\includegraphics[width=\linewidth]{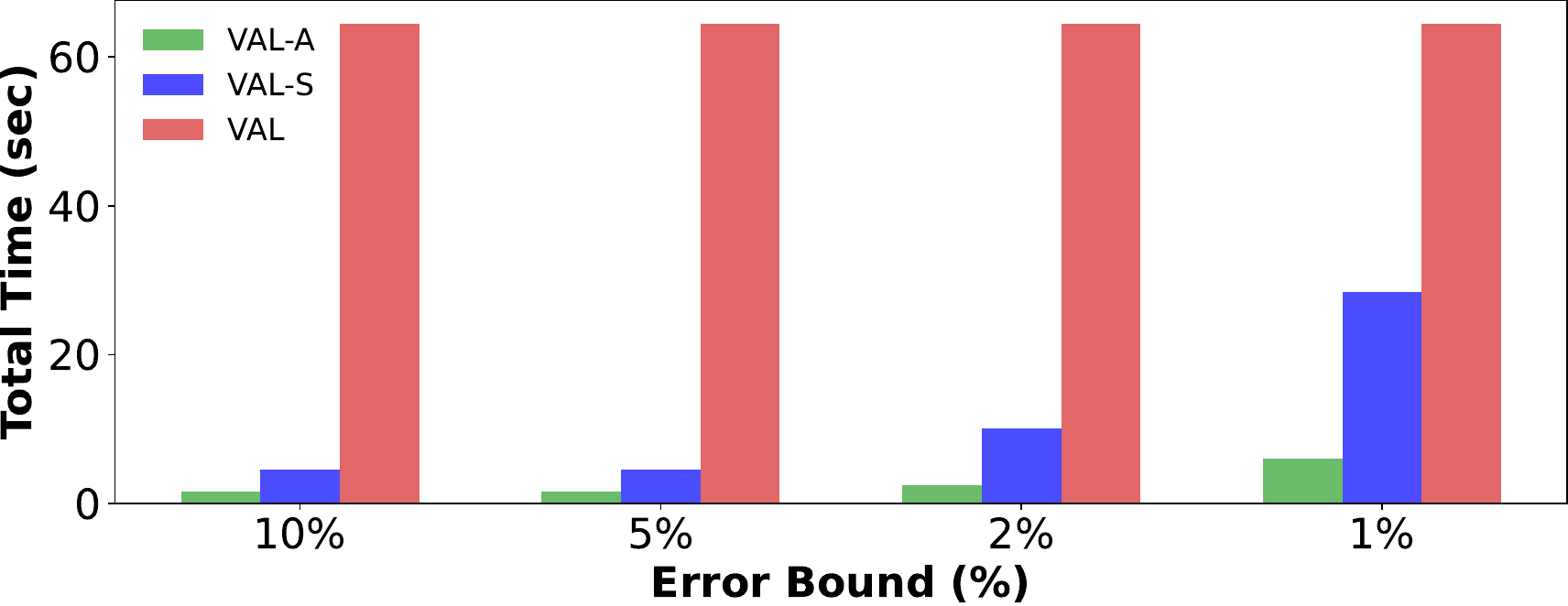}
		\caption{SYNTH50}
		\label{fig:synth50_error}
	\end{subfigure}

	\vspace{14pt}

	\begin{subfigure}{0.48\textwidth}
		\centering
		\includegraphics[width=\linewidth]{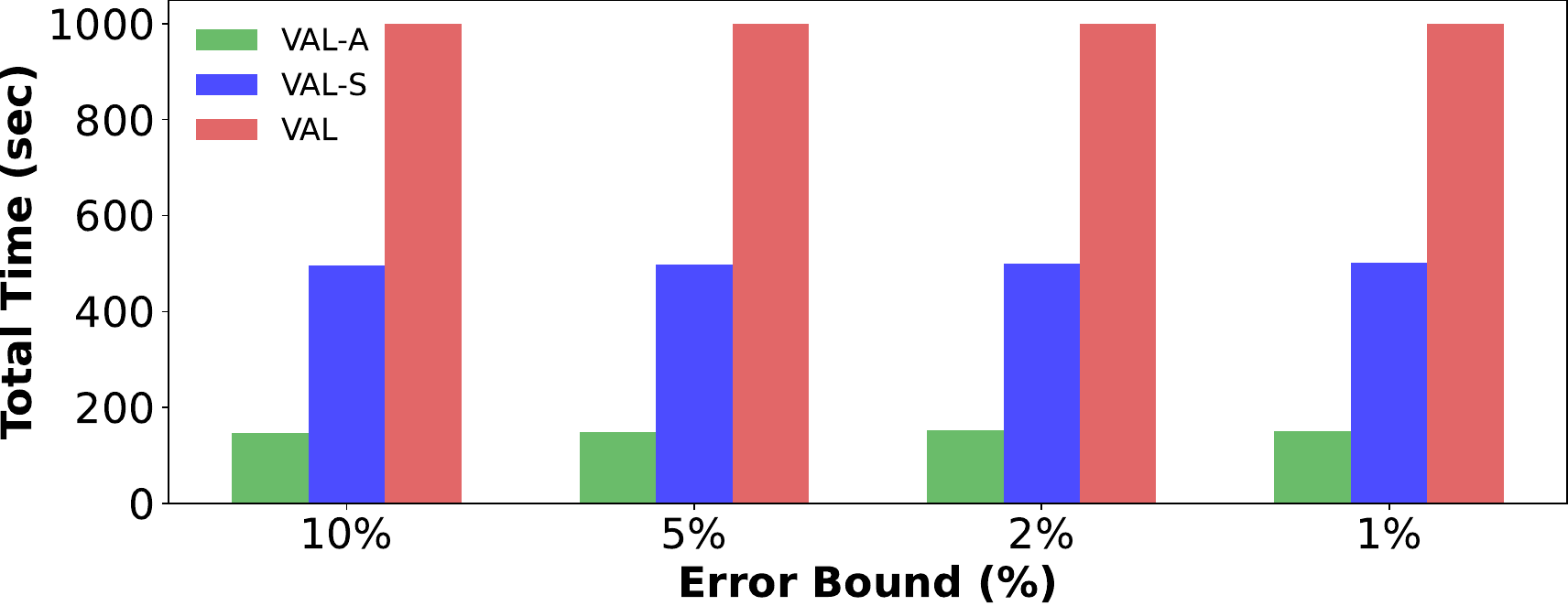}
		\caption{TAXI}
		\label{fig:taxi_error}
	\end{subfigure}

	\caption{Overall query execution time vs.\ User-defined error bound.}
	\label{fig:error_bound}
\end{figure*}

\subsection{Results}
\label{sec:res}

\subsubsection{Query Execution Time}
In this experiment, we compare the query execution time of \vala against competitors across the three datasets. Figure~\ref{pl:workload} presents the execution time for queries \( Q_1 \sim Q_{99} \), excluding the first query, which involves index initialization and construction common to all approaches. \vala and \vals are configured with a very tight $1\%$ \textit{user-defined error threshold} (\(\epsilon_{\text{max}} = 0.01\)).

Across all   datasets, \vala consistently achieves lower query execution times compared to both \val and \vals. \val, which computes exact aggregates, incurs significantly higher and more variable execution times due to the need to access all required objects for which it cannot fully utilize stored metadata from fully-contained tiles. \vals, while employing the same incremental sampling approach as \vals, does not store aggregate metadata. As a result, for every query, \vals must perform incremental sampling across its overlapping tiles, continuously retrieving object values from the raw data file until the confidence interval meets the user-defined threshold. This leads to higher query times, and for many queries, even worse performance than \mbox{\val}, which directly computes exact results. In effect,
\mbox{\vals} only utilizes the tile-based index to locate objects within the query region, sample from them, and access their   attribute values required for aggregation.

In contrast, \vala efficiently balances incremental sampling with metadata reuse, reducing execution time across the query sequence even under the strict \(1\%\) error threshold. Unlike \vals, \vala stores and updates partial metadata, leveraging previously sampled values to minimize redundant I/O operations in future queries. While both \val and \vala benefit from \textit{index adaptation}, which dynamically refines the tile structure in frequently explored areas, \val still requires reading all objects within a query region. In contrast, \vala can selectively update its stored metadata when the computed confidence interval remains within user-defined constraints, further reducing the need for full-file access.

Especially, for the {TAXI} dataset, during the last queries of the exploration scenario, the user navigates to areas with significantly fewer taxi trips. This justifies the sharp drop in execution time for all three methods, as fewer objects need to be accessed and processed.

\subsubsection{I/O Operations}
The query execution time examined above is primarily determined by the number of I/O operations required to access objects from the raw data file. This is evident in Figure~\ref{pl:io}, where the I/O plots closely follow the corresponding execution time trends from Figure~\ref{pl:workload}.
Also here, \vala and \vals are configured with a very tight $1\%$ \textit{user-defined error threshold} (\(\epsilon_{\text{max}} = 0.01\)).

For the synthetic datasets (SYNTH10/50), the I/O operation trends are nearly identical (Fig.~\ref{fig:synth10_io} \& \ref{fig:synth50_io}). This similarity is expected since both datasets contain the same number of objects with uniformly distributed attribute values in the same range. The only difference between them is the number of attributes (i.e., 10 and 50, respectively), which does not affect the number of accessed objects for answering aggregate queries.

\subsubsection{Effect of Error Bound on Performance}
 In this experiment, we evaluate how the user-defined error bound \(\epsilon_{\text{max}}\) impacts the total query evaluation time for \vala and \vals.
Figure~\ref{fig:error_bound} presents the \textit{overall query execution time} for the full sequence of queries (excluding \(Q_0\) which initializes the index) under different error bounds: 10\%, 5\%, 2\%, and 1\%.
As expected, since \val does not utilize approximate query evaluation, its performance remains constant regardless of the error bound.
\vala and \vals benefit from higher error bounds by reducing the number of samples required to meet the confidence interval constraints, resulting in lower query evaluation times.

\begin{tcolorbox}
	[colback=gray!30,colframe=white,arc=0pt,outer arc=0pt,
left=4pt, right=4pt, top=5pt, bottom=5pt]
Overall, across \emph{all datasets} and \emph{all error bounds}, \linebreak
\vala consistently outperforms both methods.
On average, \vala \textit{achieves a $3.9\times$ speedup} over \vals and $7.4\times$ over \val across all cases.
\end{tcolorbox}

\stitle{SYNTH10.} As depicted in Figure~\ref{fig:synth10_error}, \vala completes the sequence of queries in under $5$ sec even at the smallest error bound (1\%), while \vals takes around $22$ sec and \val about $32$ sec.

The total query time for \vals increases significantly as the error bound tightens from \(10\%\) to \(1\%\), reflecting the additional sampling effort required to meet the stricter confidence interval constraints. This leads to increased file accesses and higher I/O costs. \vala follows a similar trend but remains consistently faster due to its adaptive metadata reuse, which reduces the need for redundant sampling.

\stitle{SYNTH50.} In Figure~\ref{fig:synth50_error}, the performance gap increases further compared to SYNTH10: \vala completes the workload in 7 sec at 1\% error, compared to 28 sec for \vals and 64 sec for \val.

A similar pattern as in SYNTH10 is observed, but with an overall increase in query execution time due to the larger file size. The relative performance of \vala and \vals remains consistent, with \vala achieving lower total execution times.

\stitle{TAXI.}
 In  TAXI   (Fig.~\ref{fig:taxi_error}), the advantage of \vala becomes particularly clear, where the exploration area contains a high concentration of taxi trips (central Manhattan).
In this dataset, even for the tightest error bound (1\%), \vala finishes in 150 sec, while \vals takes over 500 sec and \val exceeds 1000 sec. The performance gap is particularly pronounced due to the dataset’s higher I/O overhead, making metadata reuse even more crucial.

\vala exhibits minimal variation in execution time across different error bounds. This is primarily due to the \textit{low variance} of the aggregate attribute (trip fare amount), which results in \textit{stable estimates even with small sample sizes}. As a result, the confidence interval converges quickly, and additional sampling has a diminishing impact on accuracy. In our \textit{incremental sampling approach}, we begin with an initial sampling rate  and progressively refine it based on computed confidence intervals until the required error bound is met. Since sampling stops once the confidence interval satisfies the threshold, the total number of sampled objects remains \textit{nearly the same across different error bounds}, leading to consistent query execution times. This behavior is also observed in \vals, which follows the same sampling strategy. However, \vala maintains significantly \textit{lower overall execution times} due to its \textit{metadata reuse}, which reduces redundant I/O operations.

\begin{figure}[t]
	\centering
	\includegraphics[width=1.0\linewidth]{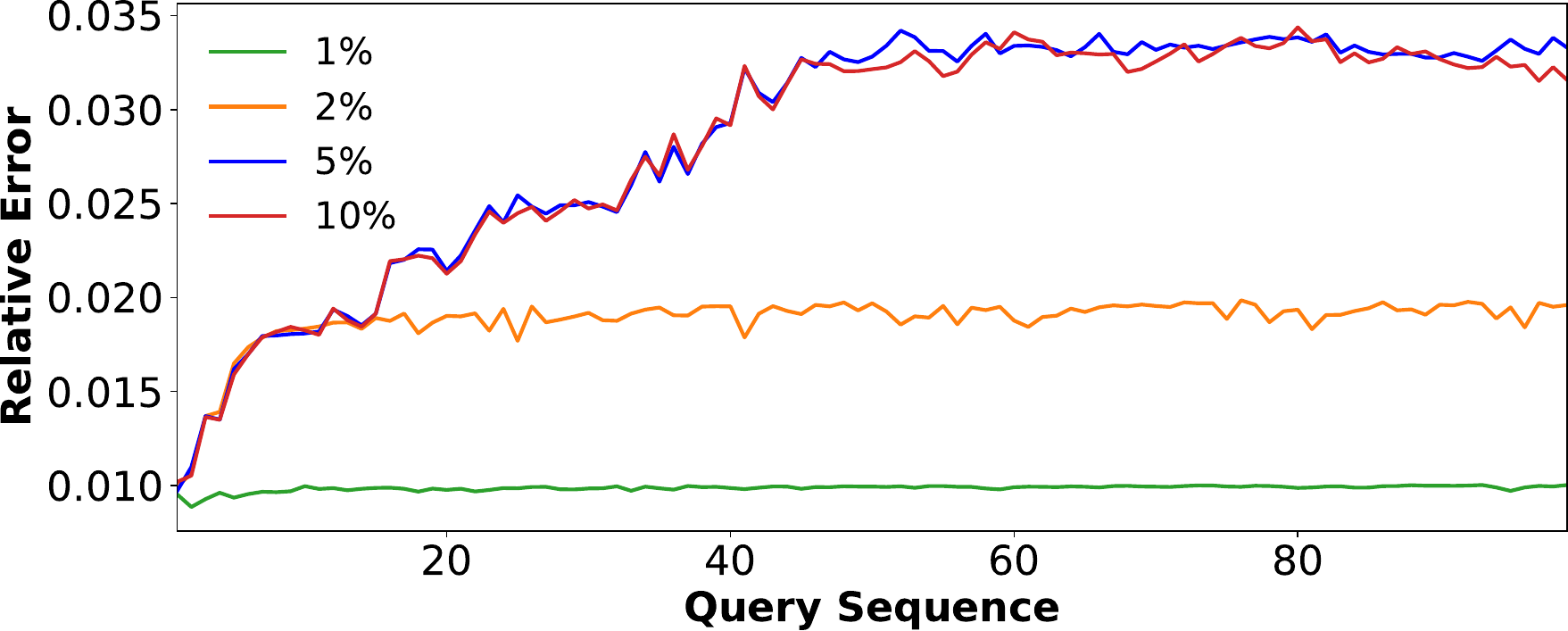}
	\caption{Relative Error for different User-defined error bounds  [SYNTH10].}
	\label{fig:error_vs_query}
\end{figure}

\subsubsection{Approximation Accuracy}
To examine how the user-defined error threshold impacts approximation accuracy, we vary the error bound \(\epsilon_{\text{max}}\) and measure the resulting relative error across the query sequence. Figure~\ref{fig:error_vs_query} presents the relative error for different thresholds (\(1\%\), \(2\%\), \(5\%\), and \(10\%\)) for \vala over the SYNTH10 dataset. Similar trends were observed for SYNTH50 and are omitted for brevity. In the TAXI dataset, due to the distribution characteristics of the aggregated attribute (taxi fare amount), different error bounds result in less pronounced variations in relative error.

As shown in the Figure~\ref{fig:error_vs_query}, the actual relative error (computed as the deviation of the estimated aggregate from the exact result, normalized by the exact value) remains \textit{below the corresponding user-defined error bound} \(\epsilon_{\text{max}}\). This confirms that \vala effectively maintains computed confidence intervals within user-specified constraints.

As expected, lower error bounds (e.g., \(1\%\)) result in consistently lower relative errors, as more samples are taken to refine estimates. Conversely, higher error bounds (\(5\%\) and \(10\%\)) lead to larger relative errors, as incremental sampling adapts dynamically based on the user-defined threshold. Interestingly, the relative error curves for 5\% and 10\% follow a similar trend. This occurs because, beyond a certain point, additional sampling has a diminishing impact on accuracy, as the confidence interval stabilizes at a similar rate for both thresholds. Consequently, both error bounds require a comparable number of samples to meet their constraints, leading to nearly identical relative error behavior. This effect is influenced by dataset-specific characteristics, such as attribute distribution.

At the start of the exploration, \vala benefits from \textit{exact metadata} stored during index initialization, as the file is parsed while constructing the index. This results in very low relative error for initial queries, as fully contained tiles in the queries contribute no uncertainty to the query result. However, as user exploration continues, the \textit{index dynamically adapts} by splitting tiles in frequently queried areas to improve granularity and maximize the number of fully contained tiles in future queries. When a tile is split, its metadata is incrementally updated during query evaluation using objects read from the file.

Since \vala utilizes \textit{incremental sampling}, it does not require full metadata storage for all tiles but instead \textit{adjusts I/O adaptively} based on the user-defined error threshold. As a result, even though relative error increases over time due to tile splits and new exploration areas, it consistently remains {\textit{well below}} the user-defined threshold \(\epsilon_{\text{max}}\). This demonstrates that \vala successfully balances \textit{efficiency and accuracy}, dynamically managing metadata storage and I/O operations while ensuring query results adhere to user-specified confidence bounds.

\subsubsection{Confidence Interval Behavior}
\begin{figure}[t]
    \centering
{
\hspace{-14pt}
    \includegraphics[width=1.0\linewidth]{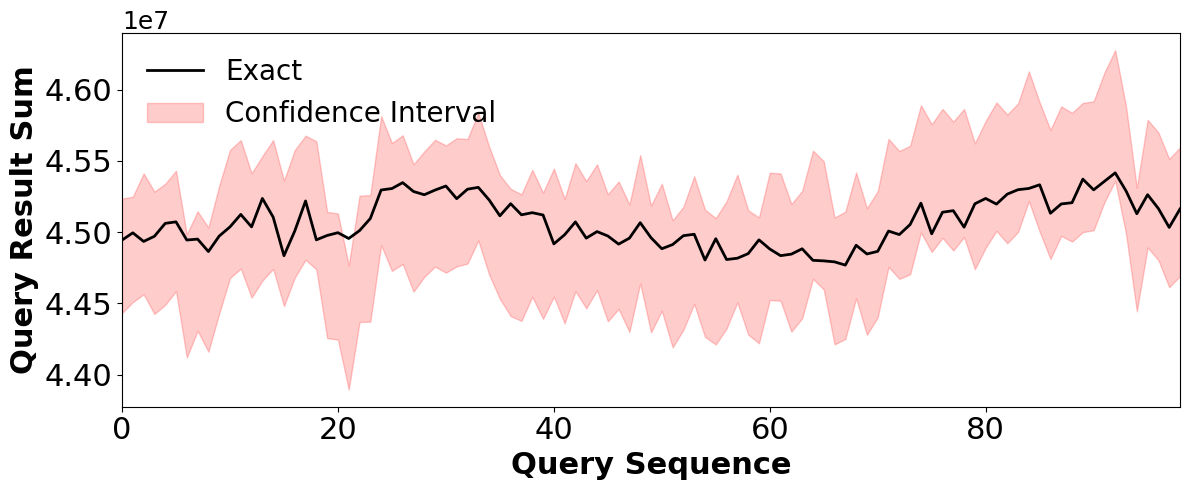}
    \caption{Exact aggregate value \& Confidence interval  [SYNTH10] (\(\epsilon_{\text{max}} = 0.01\)).}
        \label{fig:conf_interval_synth10}}
\end{figure}

To further assess approximation accuracy, we evaluate the behavior of the computed confidence intervals during the exploration scenario. Figure~\ref{fig:conf_interval_synth10} shows the confidence intervals for the estimated aggregate values (here, the sum) alongside the exact aggregate values for the SYNTH10 dataset, using a user-defined error bound of \(1\%\). Similar trends were observed for other datasets and error bounds.

As shown, the exact aggregate consistently falls within the computed confidence intervals for most queries, confirming that \vala maintains statistically valid approximations. In this experiment, we set the confidence level to \(\gamma = 0.95\), meaning that if the same query were repeated under identical conditions, the computed confidence interval would contain the exact aggregate in 95 out of 100 cases due to the randomness inherent in sampling. In practice, even when different queries are posed during exploration, the average coverage rate in our experiments is approximately 95\%.

\begin{tcolorbox}
	[colback=gray!30,colframe=white,arc=0pt,outer arc=0pt,
left=4pt, right=4pt, top=5pt, bottom=5pt]

In the results shown in  Figure~\ref{fig:conf_interval_synth10}, \textit{only 2 out of 100 queries fell outside the computed confidence interval}, confirming the reliability of our estimates.
\end{tcolorbox}

 \section{Related Work}
 \label{sec:rw}

This work is related to various fields, namely, in-situ processing, approximate processing, adaptive indexing, and visual-oriented data management.
 We briefly describe most prominent research works in these areas and justify our contribution.

 \stitle{In-situ Raw Data Processing.}
 Data loading and indexing account for a substantial portion of the time required to complete an analysis in both traditional RDBMSs and Big Data environments \cite{IdreosAJA11}. To address this bottleneck, in-situ query processing avoids loading data into a DBMS by enabling direct operations on raw data files. One of the first systems designed for querying raw data without relying on a DBMS was NoDB \cite{Alagiannis2012}, while PostgresRAW was an early adopter of in-situ query processing. PostgresRAW constructs auxiliary structures known as "positional maps" on-the-fly, which track the positions of data attributes in files and temporarily store previously accessed data in cache.
However, unlike our approach, PostgresRAW support only exact query answering.
Moreover, the  positional maps are only effective at reducing parsing and tokenization overhead during query execution, and they cannot reduce the number of objects examined in two-dimensional range queries. Additionally, our method improves the performance of aggregate queries by reducing raw file accesses through the reuse of previously calculated statistics at the tile level.

DiNoDB \cite{TianALAMV17} is a distributed version of PostgresRAW. In the same direction, PGR \cite{KarpathiotakisBAA14} extends the positional maps in order to both index and query files in formats other than CSV. In the same context, Proteus \cite{KarpathiotakisA16} supports various data models and formats.  Slalom \cite{OlmaKAAA17,slalomvldbj19} exploits the positional maps and integrates partitioning techniques that take into account user access patterns.

Raw data access methods have been also employed for the analysis of scientific data, usually stored in array-based files.
In this context, Data Vaults \cite{IvanovaKMK13} and SDS/Q \cite{BlanasWBDS14} rely on DBMS technologies to perform analysis over scientific array-based file formats.
Further, SCANRAW \cite{ChengR15} considers parallel techniques to speed up CPU intensive  processing tasks associated with raw data accesses.

RawVis \cite{IS,BikakisMPV18} exploits  VALINOR, a tile-based index in the context of in-situ visual exploration, supporting 2D visual operations over numeric attributes.
\cite{MaroulisS22f,MaroulisBPVV21,dolap} extend RawVis to support categorical-based operations, offering also a memory management mechanism.
Compared to this work, the previous versions of RawVis framework do not support approximate query answering.

Note that, several well-known DBMS support SQL querying over CSV files.
Particularly,  MySQL provides the CSV Storage Engine\footnote{\href{https://dev.mysql.com/doc/refman/8.4/en/csv-storage-engine.html}{\scriptsize https://dev.mysql.com/doc/refman/8.4/en/csv-storage-engine.html}},
Oracle offers the External Tables\footnote{\href{https://oracle-base.com/articles/12c/external-table-enhancements-12cr1}{\scriptsize https://oracle-base.com/articles/12c/external-table-enhancements-12cr1}}
and  Postgres has the Foreign Data.\footnote{\href{https://www.postgresql.org/docs/current/ddl-foreign-data.html}{\scriptsize  www.postgresql.org/docs/current/ddl-foreign-data.html}}
These tools enable interoperability with raw formats, but do not focus on user interaction, parsing the entire file for each posed query, and resulting in significantly low query performance \cite{Alagiannis2012}.

 \stitle{Approximate Processing.}
Approximate Query Processing (AQP) \cite{MozafariN15,CormodeGHJ12,LiL18,Chaudhuri17} is a long studied area offering means for rapid, “good-enough” answers in interactive data exploration.
A common thread in most of these approaches is the use of \textit{stratified sampling to deliver statistically bounded estimates} \cite{Acharya2000,AgarwalMPMMS13,Chaudhuri2001OvercomingLO,Chaudhuri2001ARO,Chaudhuri2007OptimizedSS,Sidirourgos2011SciBORQSD}.
Similar to these methods, our work leverages stratified sampling; however, we adapt the sampling strategy to the in-situ exploratory setting and on-the-fly index adaptation. Rather than relying on a pre-defined sampling ratio, our approach dynamically adjusts the sampling process based on the evolving user interactions, index adaptations and the characteristics of the  data.

 Subsequent research has explored the use of \textit{auxiliary indices and pre-aggregation techniques} to further accelerate query evaluation. Systems proposed in \cite{DBLP:conf/sigmod/DingHCC016,Moritz2017TrustBV} build auxiliary data structures that store precomputed summaries, which are then combined with sampled estimates, thereby avoiding full data scans during query execution.

 Methods based on \textit{approximate pre-aggregation} \cite{Jermaine2003RobustEW,Jin2006NewSE} compute partial aggregates either offline or on-the-fly, and then refine these estimates as additional data is processed.
 While effective in many scenarios, these techniques typically require prior data loading and pre-processing, which can be prohibitive in exploratory contexts.
 In contrast to these techniques, our method minimizes both pre-processing time and storage by initially creating a “crude” version of the index, tailored for overlapping, non-random exploratory queries. The index is then enriched as the user explores the data, thereby reducing the overhead between initialization and query execution. In essence, our approach capitalizes on the natural locality of exploratory queries to limit upfront costs while still enabling efficient, in-situ approximate query processing.

In visual data analysis, \textit{approximate processing techniques} (a.k.a.\ \textit{data reduction}), such as sampling and binning, have been widely used to improve interactivity and address \textit{information overloading}  \cite{AQP++,MaroulisVLDB24,0001S20,bsps15,KimBPIMR15,ParkCM16,liu2013immens,KwonVHD17,wickham2013bin}. Most of them follow \textit{progressive approaches} \cite{hal-04361344,FeketeF0S18,ZgraggenGCFK17,AngeliniSSS18,StolperPG14,ChenZFFW22,HograferASS22,RahmanAKBKPR17,AgarwalMPMMS13,FisherPDs12,OM323,Davos21}. Instead of performing all the computations in one step (that can take a long time to complete), they split them in a series of short chunks of approximate computations that improve with time.
Some of these works also consider visualization parameters to ensure perceptually similar visualizations and offer visualization-aware error guarantees.

More recently, techniques such as AQP++ \cite{AQP++} and PASS \cite{Liang21}
have {combined sampling-based AQP with approximate pre-aggregation} to produce
tighter confidence intervals for aggregate queries and reduce query costs.
While our method likewise leverages sampling and aggregates, it differs in
two crucial ways. First, instead of relying primarily on \emph{complete}
precomputed aggregates, we also maintain \emph{partial aggregates} from
sampling, which incrementally refine query results whenever accuracy must be improved. Second, whereas AQP++ and PASS rely on significant offline
precomputation based on predicted or fixed workloads, our approach adaptively refines sampling and index partitioning \emph{during} user exploration. Only the areas that are actually viewed and queried are progressively refined,
eliminating both up-front overhead and restrictive assumptions about future
queries. As users pan and zoom in the 2D plane, we read objects on demand from raw data files and update partial aggregates to narrow the gap between
approximate and exact results. This incremental strategy is especially
effective for {in-situ} visual analysis, where minimizing I/O costs
and handling unanticipated query regions are paramount.

This paper is influenced from our preliminary work \cite{bigvis24}, where we presented early results on adaptive indexing for approximate query answering.
In \cite{bigvis24}, we relied on exact min-max tile metadata to deterministically bound aggregates and reduce I/O. However, such bounds are often overly pessimistic, leading to wide confidence intervals and making this approach practical only when the aggregated attribute has very low variance, i.e., when its minimum and maximum values are close.
In this work, we adopt different approaches,  we develop an \textit{incremental sampling strategy} that dynamically refines query results until they meet user-defined accuracy constraints.
Additionally, we leverage sampled objects to compute and store \textit{approximate metadata}, further reducing redundant I/O.

\stitle{Indexes in Human-Data Interaction Scenarios.}
In the context of human-data interaction, several indexes have been introduced.
VisTrees \cite{El-HindiZBK16} and HETree \cite{bsps15} are tree-based main-memory
indexes that address visual exploration use cases, i.e., they offer exploration-oriented
features such as incremental index construction and adaptation.

Nanocubes  \cite{Lins2013}, Hashedcubes \cite{PahinsSSC17}, SmartCube \cite{LiuWSY20},
Gaussian Cubes \cite{WangFWBS17}, and TopKubes  \cite{Miranda2017} are main-memory data structures defined over spatial, categorical and temporal data.
The aforementioned works are based on main-memory variations of a data cube in order to reduce the time needed to generate visualizations.

Further,  graphVizdb \cite{BikakisLKG16,Bikakislkps15} is a graph-based visualization tool, which employs a  2D  spatial index (e.g., R-tree) and maps user interactions to 2D window queries.
To support scalability a partition-based graph drawing approach is proposed.
Spatial 2D indexing is also adopted in Kyrix   \cite{TaoLWBDCS19}.
Kyrix is a generic platform that supports efficient Zoom and Pan operations over arbitrary data types.
These works do not consider approximate query processing, as they require a preprocessing phase to create an index, and thus cannot be used in in-situ scenarios. Another shortcoming is that they reside in main memory, which in many cases require prohibitive amounts of memory.

In a different context, tile-based structures are used in visual exploration scenarios.
Semantic Windows \cite{KalininCZ14} considers the problem of finding rectangular regions (i.e.,
tiles) with specific aggregate properties in  exploration scenarios.
ForeCache \cite{bcs15} considers a client-server architecture in which the user visually
explores data from a DBMS. The approach proposes a middle layer, which prefetches tiles
of data based on user interaction.
Our work considers different problems compared to the aforementioned approaches.

Finally, survey papers on the broader areas of human-data interaction and visual analytics, including the involvement of AI, can be found in
\cite{WuDCLKMMSVW23,blog,WangLZ24,AndrienkoAAWR22,YuanCYLXL21,Bikakis2022,BasoleM24,RaeesMLKP24,WuWSMCZZQ22,Battle2020,YangLWL24,YeHHWXLZ24,WangCWQ22,HohmanKPC19,QinLTL20,richer2024scalability,bigvissurvey}.

 \stitle{Adaptive Indexing.}
Similarly to our work, the basic idea of approaches like database cracking and adaptive indexing is to incrementally adapt the indexes and/or refine the physical order of data, during query processing,  following the characteristics of the workload \cite{icde25upadaptive,ZardbaniMIK23,0002ZMK23,JensenLZIK21,HolandaM21,ZardbaniAK20,cidrIdreosKM07,Nerone21,HolandaM21,HolandaMMR19,NathanDAK20,PavlovicSHA18}.

However, these works neither support approximate query processing nor are they designed for the in-situ scenario.
In most cases the data has to be  previously loaded / indexed in the
  system/memory,  i.e., a preprocessing phase is considered.
Additionally, the aforementioned works refine the (physical) order of data, performing highly expensive data duplication and allocate large amount of memory resources.
Nevertheless, in the \textit{in-situ scenarios}  the analysis is performed directly over immutable raw data files considering limited resources.

Furthermore, most of the existing cracking and adaptive indexing methods have been
developed in the context of column-stores
\cite{HalimIKY12,GraefeK10,cidrIdreosKM07,IdreosMKG11,IdreosKM09,PetrakiIM15,AlexiouKL13,HolandaMMR19},
or MapReduce systems \cite{0007QSD14}.
On the other hand, our work has been developed to handle raw data stored in text files with commodity hardware.


\section{Conclusions}
\label{sec:concl}
This paper studies the problem of approximate query answering for in-situ  exploration scenarios. It focuses on 2D exploration settings, like scatter plots and maps, aiming at balancing between accuracy and performance of queries evaluated directly on large raw data files.
We propose an in-memory index, \ind, and a query processing method, which minimizes query latency by integrating adaptive indexing with approximate computation of aggregate statistics on  regions of the visited area. 
The adaptation refines index structure and approximate statistics based on user
exploration, and leverages incremental sampling to ensure accuracy while reducing redundant file accesses. 
We have evaluated our approach over real and synthetic datasets and the 
results show substantial performance gains in terms of user response times, 
compared to exact in-situ query processing techniques. 

Some interesting directions for future works involve integrating the approximate processing  methods into progressive visualization environments where query results and visualization accuracy are incrementally refined as more data becomes available or as more processing time is allocated. 
This integration will help balance index adaptation with query evaluation time, providing progressively more accurate results to the user while better adapting the index to support future operations. 
Also, introducing approximate methods not only for numerical but also for categorical attributes is challenging because the memory requirements of the index can become prohibitive as the number of categorical attributes increases. 

\bibliographystyle{IEEEtran}
\bibliography{biblio}

\end{document}